
\input harvmac

\let\preprint=0    


\def\PSfig#1#2#3#4#5{
  \nfig{#1}{#2}
  \if\preprint1
    \topinsert
      \dimen1=#3                     
      \divide\dimen1 by 100
      \multiply\dimen1 by #5
      \vskip \dimen1
      \count253=83                   
      \multiply\count253 #5
      \divide\count253 by 100
      \count254=100                  
      \advance\count254 by -#5
      \multiply\count254 by 254      
      \divide\count254 by 100
      \advance\count254 by -65       
      \dimen1=#3                     
      \advance\dimen1 by -11in
      \divide\dimen1 by 2
      \count255=\dimen1
      \divide\count255 by 100
      \multiply\count255 by \count253
      \dimen2=1bp
      \divide\count255 by \dimen2
      \advance\count255 by 18        
      \includegraphics{#4}
      \setbox1=\hbox{#1. #2}
      \ifdim \wd1 < 4.5in
         \centerline{#1. #2}
      \else
         \centerline{\vbox{\hsize 4.5in \baselineskip12pt \noindent #1. #2}}
      \fi
    \endinsert
  \fi
}


\def\refmark#1{${}^{\refs{#1}}$\ }
\def\footsym{*}\def\footsymbol{}\ftno=-2
\def\foot{\ifnum\ftno<\pageno\xdef\footsymbol{}\advance\ftno by1\relax
\ifnum\ftno=\pageno\if*\footsym\def\footsym{$^\dagger$}\else\def\footsym{*}\fi
\else\def\footsym{*}\fi\global\ftno=\pageno\fi
\xdef\footsymbol{\footsym\footsymbol}\footnote{\footsymbol}}

\font\authorfont=cmcsc10 \ifx\answ\bigans\else scaled\magstep1\fi
\edef\smalltfontsize{\ifx\answ\bigans scaled\magstep2\else scaled\magstep3\fi}
\font\smalltitlefont=cmr10 \smalltfontsize

\baselineskip=0.29in plus 2pt minus 2pt

\def\lsim{ \vcenter{\hbox{$\buildrel{\displaystyle <}\over\sim$}} }
\def\gsim{ \vcenter{\hbox{$\buildrel{\displaystyle >}\over\sim$}} }
\def\sumint{\hbox{$\sum$}\!\!\!\!\!\!\!\int}
\def\sumintf{\sumint^{\rm(f)}}

%
\def\slashmark#1#2#3{\global\setbox0=\hbox{\raise#2em
	\hbox{\kern#3em $#1\mathchar"0236$}}%
	\wd0=0pt \ht0=0pt \dp0=0pt \box0}
\def\dslash{{\mathchoice{\slashmark\displaystyle{.075}{-.075}}%
		{\slashmark\textstyle{.075}{-.075}}%
		{\slashmark\scriptstyle{.055}{-.055}}%
		{\slashmark\scriptscriptstyle{.04}{-.04}}\partial}}

\def\Lslash#1{{\mathchoice{\slashmark\displaystyle{-.1}{-.125}}%
		{\slashmark\textstyle{-.1}{-.125}}%
		{\slashmark\scriptstyle{-.075}{-.1}}%
		{\slashmark\scriptscriptstyle{-.06}{-.08}}#1}}

\def\Esp{E_{\rm sp}}
\def\Mw{M_{\rm w}}
\def\Mz{M_{\rm z}}
\def\mh{m_{\rm h}}
\def\mt{m_{\rm t}}

\def\etal{{\it et al.}}
\def\gy{g_{\rm y}}
\def\gs{g_{\rm s}}
\def\thetaw{\theta_{\rm w}}
\def\thetat{\tilde\theta}
\def\nf{n_{\rm f}}
\def\d{{\rm d}}
\def\const{{\rm const.}}
\def\Tc{T_{\rm c}}
\def\meff{m_{\rm eff}}
\def\phic{\phi_{\rm c}}
\def\ML{M_{\rm L}}
\def\MLw{M_{\rm Lw}}
\def\MLz{M_{\rm Lz}}
\def\MLg{M_{\rm L\gamma}}
\def\MLa{M_{\rm L1}}
\def\MLb{M_{\rm L2}}
\def\MLc{M_{\rm L3}}
\def\Mdebye{M_{\scriptscriptstyle\rm debye}}
\def\dMa{\Delta{\cal M}_{\rm 2}}
\def\dMb{\Delta{\cal M}_{\rm 1}}
\def\mf{m_{\rm f}}
\def\mfa{m_{\rm f2}}
\def\mfb{m_{\rm f1}}
\def\gammaE{\gamma_{\rm\scriptscriptstyle E}}
\def\eps{\epsilon}
\def\MSbar{$\overline{\rm MS}$}
\def\intk{\int{\d^{3-2\eps}k\over(2\pi)^{3-2\eps}}}
\def\nsum{\sum_{n=1}^\infty}
\def\lsum{\sum_{l=0}^\infty}
\def\eg{{\it e.g.}}
\def\etc{{\it etc.}}
\def\intpthree{\int {\d^3 p\over(2\pi)^3}}
\def\intkthree{\int {\d^3 k\over(2\pi)^3}}
\def\intqthree{\int {\d^3 q\over(2\pi)^3}}
\def\Jf{J_{\rm f}}
\def\If{I_{\rm f}}
\def\Hff{H_{\rm ff}}

\def\Vcl{V_{\rm cl}}
\def\Veff{V_{\rm eff}}
\def\Vone{V^{(1)}}
\def\Vring{\Vone_{\rm ring}}
\def\naive{{\hbox{\rlap{$\rm\scriptstyle R$}/}}}
\def\bare{{\rm bare}}

\def\mmu{\nu}      
\def\mmueff{\mmu_{\rm eff}}
\def\mub{\bar\mu}  

\def\cH{c_{\rm\scriptscriptstyle H}}
\def\cb{c_{\rm\scriptscriptstyle B}}      
\def\cf{c_{\rm\scriptscriptstyle F}}
\def\ceps{\iota_\eps}
\def\cepsf{\iota_{\eps}^{\rm(f)}}

\def\Jepslog{ {1\over\eps} + \ln\left(\mub^2\over T^2\right) - 2\cb }
\def\Jepslogf{ {1\over\eps} + \ln\left(\mub^2\over T^2\right) - 2\cf }
\def\Jepslogc{ {1\over\eps} + \ceps
                           + \ln\left(\mub^2\over T^2\right) - 2\cb }

\lref\linde{
  D.A. Kirzhnits and A.D. Linde,
    Phys.\ Lett.\ \bf D9\rm, 2257 (1974);
    Ann.\ Phys.\ \bf 101\rm, 195 (1976).
}
\lref\weinberg{S. Weinberg, Phys.\ Rev.\ \bf D9\rm, 2257 (1974).}
\lref\dolan{L. Dolan and R. Jackiw, Phys.\ Rev.\ \bf D9\rm, 3320 (1974).}
\lref\dolanb{L. Dolan and R. Jackiw, Phys.\ Rev.\ \bf D9\rm, 1686 (1974).}
\lref\takahashi{K. Takahashi, Z. Phys.\ \bf C26\rm, 601 (1985).  The
  reader should beware that the gauge theory cases are handled incorrectly
  in this paper.}
\lref\fendley{P. Fendley, Phys.\ Lett.\ \bf B196\rm, 175 (1987).}
\lref\carrington{M. Carrington, Phys.\ Rev.\ \bf D45\rm, 2933 (1992).}
\lref\parwani{R. Parwani, Phys.\ Rev. \bf D45\rm, 4695 (1992).}
\lref\nielsen{For a discussion of the gauge dependence of the effective
  potential at {\it zero} temperature, see
  N.K. Nielsen, Nucl.\ Phys.\ \bf B101\rm, 173 (1975);
  I.J.R. Aitchison and C.M. Fraser, Ann.\ Phys.\ \bf 146\rm, 1 (1984).}
\lref\wwu{E. Weinberg and A. Wu, Phys.\ Rev. \bf D36\rm, 2474 (1987).}
\lref\ginsparg{
  P. Ginsparg, Nucl.\ Phys.\ \bf B170\rm\ [FS1], 388 (1980).
}
\lref\reviews{For reviews, try D.J. Gross, R.D. Pisarski and L.G. Yaffe,
  Rev.\ Mod.\ Phys.\ \bf 53\rm, 43 (1981);
  J.I. Kapusta, \sl Finite-temperature field theory \rm
  (Cambridge Univ. Press: Cambridge, 1989).
}
\lref\arnold{P. Arnold, \sl Phys. Rev. \bf D46\rm, 2628 (1992).}
\lref\shaposhnikov{
  M. Shaposhnikov, JETP Lett.\ \bf 44\rm, 405 (1986);
    Nucl.\ Phys.\ \bf B299\rm, 707 (1988),
}
\lref\mclerrana{
  L. McLerran, Phys.\ Rev.\ Lett.\ \bf 62\rm, 1075 (1989).}
\lref\tzmsv{
  N. Turok and J. Zadrozny, Phys.\ Rev.\ Lett.\ \bf 65\rm, 2331 (1990);
    Nucl.\ Phys.\ \bf B358\rm, 471 (1991);
  L. McLerran, M. Shaposhnikov, N. Turok and M. Voloshin,
    Phys.\ Lett.\ \bf 256B\rm, 451 (1991).
}
\lref\cohen{
  A. Cohen, D. Kaplan and A. Nelson,
    Phys.\ Lett.\ \bf B245\rm, 561 (1990);
    Nucl.\ Phys.\ \bf B349\rm, 727 (1991);
    Phys.\ Lett.\ \bf B263\rm, 86 (1992).
}
\lref\dine{
  M. Dine, P. Huet, R. Singleton and L. Susskind,
    Phys.\ Lett.\ \bf B257\rm, 351 (1992);
  M. Dine, P. Huet and R. Singleton, Nucl.\ Phys.\ \bf B375\rm, 625 (1992).
}
\lref\sakharov{A. Sakharov, Pisma ZhETF \bf 5\rm, 32 (1967).}
\lref\manton{
  F. Klinkhammer and N. Manton, Phys.\ Rev.\ \bf D30\rm, 2212 (1984)
}
\lref\krs{
  V. Kuzmin, V. Rubakov and M. Shaposhnikov,
    Phys.\ Lett.\ \bf B308\rm, 885 (1988).
}
\lref\arnolda{
  P. Arnold and L. McLerran, Phys.\ Rev.\ \bf D37\rm, 1020 (1988).
}
\lref\khlebnikov{
  S. Khlebnikov and M. Shaposhnikov, Nucl.\ Phys. \bf B308\rm, 885 (1988).
}
\lref\shapMSM{
  M. Shaposhnikov, Phys.\ Lett. \bf B277\rm, 324 (1992);
  {\it ibid.}\ \bf B282\rm, 483(E) (1992).
}
\lref\dinebound{
  M. Dine, R. Leigh, P. Huet, A. Linde and D. Linde,
    Phys.\ Lett.\ \bf B283\rm, 319 (1992);
    Phys.\ Rev.\ \bf D46\rm, 550 (1992).
}
\lref\pdb{
  Particle Data Group, Review of Particle Properties
  in Phys. Rev. D45, number 11-II (1992).
}
\lref\boyd{
  C.G. Boyd, D. Brahm and S. Hsu, Chicago U. EFI preprint EFI-92-22 (1992).
}
\lref\chanowitz{
  M. Chanowitz, M. Furman and I. Hinchliffe,
    Nucl.\ Phys.\ \bf B159\rm, 225 (1979).
}
\lref\lep{
  T. Mori, ``Searches for the neutral Higgs Boson at LEP,'' Talk
  given at the XXVI Int Conf on HEP, Dallas, Texas, 6-12 August,
  1992.
}
\lref\magmass{
  P. Arnold, in preparation.
}

\def\apndxResults{A}
\def\apndxFermion{B}
\def\apndxCriticism{C}

\Title{\vbox{
    \hbox{UW/PT-92-18}
    \hbox{USM-TH-60}
  }}{\vbox{
    \smalltitlefont
    \centerline{The Effective Potential and First-Order Phase Transitions:}
    \centerline{Beyond Leading-Order}
  }}
\centerline{\authorfont Peter Arnold}
\centerline{\sl Department of Physics, University of Washington,
    Seattle, WA  98195}
\bigskip
\centerline{\authorfont Olivier Espinosa}
\centerline{\sl Department of Physics, University of Washington,
    Seattle, WA  98195}
\centerline{\sl Dept.\ de F{\i}sica, Univ.\ T\'ecnica Federico Santa Mar{\i}a,
    Casilla 110-V, Valpara{\i}so, Chile\foot{permanent address}}

\vskip .1in
Scenarios for electroweak baryogenesis require an understanding of
the effective potential at finite temperature near a first-order electroweak
phase transition.  Working in Landau gauge, we present a calculation of
the dominant two-loop corrections to the ring-improved one-loop potential
in the formal limit $g^4 \ll \lambda \ll g^2$, where $\lambda$
is the Higgs self-coupling and $g$ is the electroweak coupling.
The limit $\lambda \ll g^2$ ensures that the phase transition is
significantly first-order, and the limit $g^4 \ll \lambda$ allows
us to use high-temperature expansions.  We find corrections from 20 to
40\% at Higgs masses relevant to the bound computed for baryogenesis
in the Minimal Standard Model.
Though our numerical results seem to still rule out Minimal Standard
Model baryogenesis, this conclusion is not airtight because the
loop expansion is only marginal when corrections are as big as 40\%.
We also discuss why super-daisy approximations do not correctly
compute these corrections.

\Date{December 1992}

\newsec{Introduction}

Recent scenarios\refmark{\shaposhnikov\mclerrana\tzmsv\cohen-\dine}
for baryogenesis via the baryon number anomaly of the
Standard Model have stimulated a flurry of investigation into the details
of the electroweak phase transition in the hot, early Universe.
Sakharov's generic conditions for baryogenesis\refmark\sakharov\
require (1) baryon number
violation, (2) CP violation, and (3) disequilibrium.
In recent scenarios, the source of baryon number violation is
a Standard Model effect, arising because baryon number has an electroweak
anomaly and so is violated by nonperturbative physics.  The rate of
such violation is of order $\exp(-\Esp(T)/T)$ where $T$ is the temperature,
$\Esp(T) \sim \pi \sigma(T)/g_w$ is the electroweak sphaleron mass, and
$\sigma(T)$ is the Higgs vacuum expectation value (VEV) at
high temperature.\refmark{\manton\krs\arnolda-\khlebnikov}
In almost all scenarios, disequilibrium is provided by bubble
nucleation and expansion during a first-order electroweak phase
transition.
The requirement that the first-order phase
transition be sufficiently strong then places constraints on models of
the Higgs sector.  In particular, it is necessary that baryon-number
violation be turned off after the phase transition is completed; otherwise,
the Universe will simply relax back to equilibrium, where the net baryon
number is zero.  Since the rate of baryon-number violation is
exponentially sensitive to the VEV $\sigma(T)$, it is important to
study the minimum $\phi=\sigma(T)$ of the finite-temperature effective
potential $V(\phi,T)$ just after the phase transition.
The purpose of this paper is to study the extent to which two-loop
effects modify the results that have been found using one-loop
approximations to the effective potential.

In this paper, for the sake of simplicity, we shall focus on the Minimal
Standard Model with a single Higgs doublet.  Many of our calculations
should be easily extendable to more complicated Higgs sectors.
Most scenarios for electroweak baryogenesis are based on multi-Higgs
models because the CP violation required for baryogenesis can be
incorporated by CP violating interactions in the Higgs sector,
where it directly affects classical processes involving the sphaleron.
It has been suggested, however, that the standard CP violation of the quark
sector may by itself provide sufficient CP violation, so that even the
the Minimal Standard Model, with a single Higgs doublet, could be
viable.\refmark\shapMSM\
In the Minimal Standard Model, and using the one-loop ring-improved
effective potential (which we review later),
Dine \etal\refmark\dinebound\ have
established an upper bound of roughly 30 to 40 GeV on the Higgs boson
mass $\mh(0)$ (measured at zero temperature) if baryogenesis is to occur
at the weak phase transition.
This scenario is then excluded by the experimental bound of $\mh(0) > 60$
GeV.\refmark\lep\
(These particular constraints may be evaded by models
with more than one Higgs boson.)

An important question for such limits is whether they are significantly
modified by higher-order corrections.  As we shall review in the next
section, the loop expansion parameter in this context turns out to be
formally
of order $\lambda/g^2$, or equivalently $\mh(0)/\Mw(0)$, where
$\lambda$ is the Higgs self-coupling and $g$ is the SU(2) gauge
coupling.  For Dine \etal's upper bound on $\mh(0)$ of 30 to 40 GeV,
it is not {\it a priori} clear whether $\mh(0)/\Mw(0)$ is small---it
all depends on the exact numerical coefficients in the loop expansion.
Indeed, some of the earliest studies of corrections to the
one-loop potential found large corrections,
but these studies were subsequently shown to be flawed.\refmark\dinebound\
To resolve whether the bounds are reliable, we present an explicit
computation of the dominant two-loop corrections to the one-loop
ring-improved effective potential.  Our objective is to compute
the dominant corrections as the Higgs mass gets large, but not
so large that the loop expansion breaks down.  Specifically, we
assume
\eqn\orders{
  g^4 \ll \lambda \ll g^2.
}
(As we shall review in the next
section, the lower limit $g^4 \ll \lambda$ simplifies the calculation
by justifying a high-temperature expansion.)
{}From the numerical results of this calculation, we shall determine the
significance of multi-loop corrections as a function of Higgs mass;
when we find numerical
corrections of order 100\% or more for a given Higgs mass, we shall know that
the loop expansion has indeed broken down.

The computation of such corrections
has been previously attempted in the literature using a super-daisy
approximation.\refmark\boyd\
Unfortunately, as we discuss in Appendix \apndxCriticism, the
super-daisy approximation does not correctly compute the dominant
corrections to the ring-improved one-loop potential.

Naively, we expect that corrections to the baryogenesis bounds should
be small because the sphaleron is a classical solution of the
effective high-temperature theory.  The actions of classical
solutions (in this case corresponding to the Boltzmann exponent
$\Esp(T)/T$ for baryon violation) are generally inversely
related to loop expansion parameters.  So in any case where
baryon violation is turned off after the phase transition,
we should expect that the loop expansion will be under control.
Our results only marginally validate this conclusion.

The casual reader who is interested only in our final numerical results
should turn to figures 25---28 in section 8.

In section 2, we review the one-loop ring-improved potential and discuss
how to power-count diagrams to find the dominant 2-loop corrections
for $g^4 \ll \lambda \ll g^2$.  We will be led to adopt the
formal power-counting rule $\lambda\sim g^3$ for the rest of the
paper.  In section 3, we warm up to the problem of 2-loop thermal
calculations by first concocting a scalar problem that is loosely
analogous.  We shall review the equivalence of various prescriptions
for resummation of ring diagrams and we shall settle on one that
implements resummation only for the effective three-dimensional
theory obtained after heavy, non-static modes have been integrated
out.  In section 4, we turn to the simplest gauge theory case---the
Abelian Higgs model.  As well as computing the 2-loop potential near
the phase transition, we isolate which contributions are important
for shifting the VEV at the phase transition from its one-loop value.
Section 5 is devoted to the contributions of chirally-coupled fermions.
In section 6 we discuss non-Abelian theories using SU(2) as an example,
and we turn
to the Minimal Standard Model in particular in section 7.
Numerical results for the size of 2-loop corrections in
the Minimal Standard Model are presented in section 8.
Most of the results for diagrams contributing to the 2-loop potential
are collected in Appendix A.  Appendix B gives derivations of the
high-temperature expansions of some quantities discussed in the main
text.  Finally, Appendix C contains our criticism of the super-daisy
approximation to the effective potential.

Throughout this paper we shall find it convenient to work
exclusively in the Euclidean (imaginary time) formulation of thermal
field theory.  We shall conventionally refer to 4-momenta with
capital letters $K$ and to their components with lower-case letters
$K = (k_0,\vec k)$.  All 4-momenta are Euclidean, with discrete
frequencies $k_0 = 2n \pi T$ for bosons and $(2n+1)\pi T$ for fermions,
unless stated otherwise.

\newsec{Power Counting and Review}

\subsec{Pure scalar theory}

The classical, zero-temperature Higgs potential is of the form
\eqn\higgsCl{
  \Vcl(\phi) = - {1\over2} \mmu^2 \phi^2 + {1\over4!} \lambda \phi^4 .
}
For simplicity, let us temporarily ignore the gauge and fermion sectors
and review the effect of finite temperature in a theory with a single,
real scalar field.
At high temperature $T$ ($T\gg\mmu$), there is an additional
contribution to the scalar mass; the effective potential is approximately
of the form
\eqn\ShiggsT{
  \Veff(\phi,T) \approx
           {1\over2} \left(- \mmu^2 + {1\over24} \lambda T^2\right) \phi^2
         + {1\over4!} \lambda \phi^4 ,
}
which is the same as the zero-temperature potential except that
\eqn\Smueff{
  \mmu^2 \rightarrow \mmueff^2 = - {1\over24} \lambda T^2 + \mmu^2.
}
The addition of the thermal mass term above is responsible for
symmetry restoration at high temperature,\refmark{\linde\weinberg-\dolan}
and this approximation to
the effective potential describes a second-order phase transition
at $\Tc^2 \approx 24 \mmu^2/\lambda$.

\PSfig\figa{
  Quadratically divergent loop giving rise to the thermal mass.
}{8cm}{fig1.psx}{60}

Diagrammatically, the thermal mass term in \ShiggsT\ arises from
the quadratically divergent loop of \figa.
In the temperature-dependent piece of a quadratically divergent loop,
the UV divergence is cut off at momenta of order $T$, and so \figa\ is of
order $\lambda T^2$ for temperatures large compared to the scalar mass.

The thermal mass above is simply the dominant term in a high-temperature
expansion of the full finite-temperature 1-loop potential.
The temperature-dependent piece of the full 1-loop potential is simply
the free energy of an ideal gas of scalar particles with mass
$m^2(\phi) = \Vcl''(\phi)$, which is the mass associated with fluctuations
in the scalar field around a background field $\phi$:\refmark\dolan
\eqn\SVoneLoop{
  \eqalign{
    \Vone(\phi,T) = \Vcl(\phi)
      &+ T \int {\d^3 k\over(2\pi)^3}
           \ln\left\{1 -
              \exp\left[ -\beta\sqrt{k^2+m_i^2(\phi)}\right]\right\}
  \cr
      &+ (\hbox{\rm 1-loop zero-temperature result}) .
  }
}
In the high-temperature limit $T \gg m(\phi)$,
\eqna\SVone
$$
  \eqalignno{
    \Vone(\phi,T) &= \Vcl(\phi) +
        \const + {1\over24} m^2(\phi) \, T^2
               - {1\over12\pi}m^3(\phi) \, T
               + O(m^4)
  &\SVone a\cr
    &=  \const + {1\over2} \left(-\mmu^2 + {1\over24} \lambda T^2\right) \phi^2
               - {1\over12\pi} \left( -\mmu^2
                                     + {1\over2}\lambda\phi^2\right)^{3/2} T
               + {1\over4!} \lambda \phi^4
  \cr
    & \phantom{= \const}
        + O(m^4) .
  &\SVone b\cr
  }
$$
The constants above are temperature dependent but $\phi$ independent.
Such constants are not relevant to studying the mechanics of the phase
transition, and we shall generally ignore them.

\PSfig\figb{
  A generic example of a one-loop ring (daisy) graph.
}{14cm}{fig2.psx}{60}

\PSfig\figc{
  Resummation of the propagator in the ring approximation.
}{3cm}{fig3.psx}{90}

We have used the classical relation
$m^2(\phi) = -\mmu^2 + {1\over2}\lambda\phi^2$
above;
however, the effective value \Smueff\ of $\mmu^2$ is quite different
from its classical value at high temperature ($T \gsim \Tc$).
It is therefore important to make the replacement \Smueff\ and use
instead
\eqn\Smeff{
  \meff^2(\phi) = -\mmu^2 + {1\over24}\lambda T^2
                         + {1\over2}\lambda\phi^2 ,
}
which inserted into \SVone{a}\ yields the ring-improved one-loop
potential\refmark{\takahashi\fendley-\carrington}\foot{
  This potential has an imaginary part at small $\phi$.
  A physical interpretation may be found in ref.~\wwu.
}
\eqn\SVring{
  \eqalign{
    \Vring(\phi,T)
     =  \const &
              + {1\over2} \left(-\mmu^2 + {1\over24} \lambda T^2\right) \phi^2
              - {1\over12\pi} \left( -\mmu^2 + {1\over24}\lambda T^2
                                       + {1\over2}\lambda\phi^2\right)^{3/2} T
  \cr&
              + {1\over4!} \lambda \phi^4
              + O(\meff^4) .
  }
}
This potential sums the dominant contributions of the one-loop ring
(or daisy) graphs shown in \figb,
where each quadratically divergent
ring has been approximated by its high-temperature limit
$\lambda T^2/24$.  The substitution of $\mmueff^2$ for $\mmu^2$
corresponds to a resummation of the propagator as in \figc.
We shall treat this resummation more carefully and systematically
in section 3.

\PSfig\figd{
  Qualitative form of the potential as a function of temperature during
  a first-order phase transition.  The potentials have been normalized
  to be zero at the origin.
}{10cm}{fig4.psx}{70}

The potential \SVring\ {\it appears} to describe a first-order phase
transition, as shown in \figd.
At the temperature $T_0$ where the
quadratic term vanishes, the potential is of
the form $-b\lambda^{3/2}\phi^3 T + c\lambda\phi^4$ and has a minimum at
non-zero $\phi \sim \sqrt\lambda T$.
Just slightly above $T_0$, the small quadratic term will generate a second
minimum at $\phi=0$, indicating a first-order transition.
The symmetry-breaking minimum at the phase transition,
labeled $\phic$ in \figd, will occur at a point where all three
terms of the potential \SVring\ are the same order of magnitude.
One easily concludes that
\eqn\Sphic{
   \phic \sim \sqrt\lambda T,
   \qquad
   -\mmueff^2(\Tc) \sim \lambda^2 T^2 .
}

In fact, the phase transition in the pure scalar theory is known to be
{\it second} order.\foot{
  See any textbook on critical phenomena.  Concerning the order of the
  transition in the gauged case, see Ref.~\ginsparg\ for an analysis
  more generally applicable than what we shall review below.
}
The conclusions from the previous paragraph cannot be
trusted because higher-loop corrections to the 1-loop ring-improved potential
are large at $\phic$.  As discussed earlier, loops with quadratic
UV divergences are of order $\lambda T^2$ and are dominated by
large momenta of order $T$.
These $O(\lambda T^2)$ contributions are all absorbed by using
ring-improved propagators.
UV convergent loops (and the non-divergent pieces of
quadratically divergent ones) are instead
dominated by their infrared behavior.
In Euclidean space, this means loop momenta are dominated by
$k_0 = 0$ and $|\vec k| \sim m$.  The dominant $k_0=0$ piece of
the finite-temperature frequency sum $T\sum\nolimits_{k_0}$ gives
such loops a linear, rather than quadratic, dependence on T.\foot{
  An early discussion of this power counting may be found in ref.~\weinberg.
}
Including a factor of $1/\meff$ to make a dimensionless quantity,
the cost of each loop is therefore $\lambda T / \meff$.\foot{
  Loops which are logarithmically UV divergent may cost a factor of
  $(\lambda T/\meff) \ln(T/\meff)$.  We shall ignore the logarithms
  when power counting.
}
Now consider the loop expansion parameter $\lambda T / \meff$
at the minimum $\phic \sim \sqrt\lambda T$ corresponding to the
apparent first-order phase transition of the ring-improved one-loop
potential \SVring.  Eqn.~\Sphic\ implies that $\meff(\phic) \sim \lambda T$,
and so the loop expansion parameter is $O(1)$, verifying that the
ring-improved loop expansion cannot be trusted to distinguish
between a first and second-order phase transition in this model.

\subsec{Gauge theories}

The situation is quite different when the gauge sector is included.
As we now review, the first-order phase transition seemingly
described by the ring-improved effective potential is associated
with a {\it small} loop expansion parameter if the Higgs mass is
sufficiently small.  The phase transition is therefore indeed first order,
and the ring-improved loop expansion is a valid tool for studying it.

\PSfig\fige{
  Dominant contributions to the scalar thermal mass in gauge
  theories with fermions, in addition to \figa.
}{5.5cm}{fig5.psx}{90}

As in the pure scalar theory, there is a thermal contribution to the
Higgs mass.
It is of the form
\eqn\Gmueff{
  -\mmu^2 \rightarrow -\mmueff^2 = -\mmu^2 + a g^2 T^2 .
}
where $a g^2$ symbolizes a linear combination of the squared couplings
in the theory.  In the Minimal Standard Model, for instance,
\eqn\agMSM{
  a g^2 \rightarrow \left( \lambda + {9\over4} g_2^2 + {3\over4} g_1^2
                            + 3 \gy^2 \right) {T^2\over12} ,
}
where $g_2$, $g_1$, and $\gy$ are the couplings for SU(2), U(1), and the
top quark Yukawa interaction.  (We shall always ignore all Yukawa couplings
except for the top quark.  See section 7 for coupling
normalizations.)  Diagrammatically, these contributions again arise from
quadratically divergent loops, such as in \fige.

The temperature-dependence of the full finite-temperature 1-loop potential
may again be interpreted in terms of the free energy of an ideal gas.
Now we must sum contributions from all the various particles in the
theory with masses $m_i(\phi)$ induced by the background Higgs field
$\phi$:
\eqn\VoneLoop{
  \eqalign{
    \Vone(\phi,T) = \Vcl(\phi)
      &+ \sum_i \pm n_i T \int {\d^3 k\over(2\pi)^3}
           \ln\left\{1 \mp
              \exp\left[ -\beta\sqrt{k^2+m_i^2(\phi)}\right]\right\}
  \cr
      &+ (\hbox{\rm 1-loop zero-temperature result})
  \cr
    = \Vcl(\phi)
      &+ \sum_i n_i \, \Delta V_i(\phi,T)
   }
}
where the sum is over all particle species $i$, $n_i$ is the number of
degrees of freedom associated with each species, and the upper (lower)
sign is for bosons (fermions).
The high-temperature limit $T >> m_i(\phi)$ is\refmark\dolan
\eqn\dVhighT{
  \eqalign{
    &\Delta V_i(\phi,T) =
        \const + {1\over24} m_i^2(\phi) \, T^2
               - {1\over12\pi}m_i^3(\phi) \, T
               + O(m_i^4),
    \quad {\rm (bosons)}
  \cr
    &\Delta V_i(\phi,T) =
        \const + {1\over48} m_i^2(\phi) \, T^2
               \phantom{~ - {1\over12\pi}m_i^3(\phi) \, T }
               + O(m_i^4).
    \quad {\rm (fermions)}
  }
}
As before, we shall generally ignore the $\phi$-dependent constants
indicated by ``$\const$'' above.

For gauge bosons and the top quark, $m_i(\phi)$ is proportional to
$g\phi$ where $g$ is the appropriate coupling.  Ignoring the Higgs
contribution for now, the high-temperature expansion of the one-loop potential
is then schematically of the form
\eqn\Vcubic{
  \Vone(\phi,T) \approx {1\over2} (-\mmu^2 + a g^2 T^2) \phi^2
                    - b g^3 \phi^3 T
                    + {1\over4!} \lambda \phi^4 ,
}
where for the Standard Model
\eqn\bMSM{
   b g^3 \phi^3 T \rightarrow
   \left[ 6 \Mw^3(\phi) + 3 \Mz^3(\phi) \right] {T\over12\pi} =
   \left[ {3\over4}g_2^3 + {3\over8}(g_1^2+g_2^2)^{3/2} \right]
                      \phi^3 {T\over12\pi} .
}
Because of the $\phi^3$ term, the potential \Vcubic\ describes a first-order
phase-transition.

\PSfig\figf{
  Generic one-loop ring diagram in a gauge theory with fermions.
}{14.5cm}{fig6.psx}{60}

As before, we can ring-improve the one-loop potential to sum one-loop
ring diagrams such as \figf.
In Euclidean space, the small, hard
loops are quadratically divergent and contribute a thermal mass of order
$g^2 T^2$ to the $A_0$ polarization of the gauge fields.   This is the
usual Debye mass, and the ring improvement is implemented by incorporating
it into $M(\phi)$ for that polarization.\refmark\carrington\
For instance, the W contribution
to the cubic term \bMSM\ changes from
\eqn\bMSMr{
    6 \Mw^3(\phi) {T^2\over12\pi}
    \qquad {\rm to} \qquad
    \left[ 4 \Mw^3(\phi) + 2 \MLw^3(\phi) \right] {T^2\over12\pi} ,
}
where
\eqn\Mdef{
    \Mw^2(\phi) = {1\over4} g_2^2 \phi^2 ,
    \qquad
    \MLw^2(\phi) = {1\over4} g_2^2 \phi^2 + {11\over6} g_2^2 T^2 .
}
For Euclidean frequency $k_0 = 0$, which dominates the infra-red (IR)
behavior of
loops, we shall refer to the $A_0$ polarization as the {\it longitudinal}
polarization and the other two polarization perpendicular to the
four-momentum as the transverse polarizations.  The ring-improvement of
the terms arising from the Z are more complicated due to mixing with the
photon,\refmark\carrington\
and we leave explicit formulas for section 7.

We now estimate the order of magnitude of parameters associated with
the phase transition.  Consider the schematic form of the potential
\Vcubic.  (The ring improvement will not modify the following
order-of-magnitude estimates.)  As in the scalar case, the non-zero
minimum at the phase transition occurs when all three terms are roughly
the same order of magnitude, which yields
\eqn\Gorder{
   \phic \sim {g^3\over\lambda} T,
   \qquad
   -\mmueff^2 \sim {g^6\over\lambda} T^2 .
}
Table 1 shows the order of magnitude of several other parameters.
Of particular interest is the loop expansion parameter for loops
involving massive gauge bosons; it is order $\lambda/g^2$.  We
shall formally assume $\lambda \ll g^2$ (that is, $\mh \ll \Mw$ at
zero temperature) so that the loop expansion is well-behaved.
(In section 3, we shall discuss in more detail why the {\it vector}
loop expansion parameter is the relevant one for our calculation.)
Furthermore,
to justify our ubiquitous
high-temperature expansion $T \gg M$ and $m$, Table 1 shows
that we must also assume $g^4 \ll \lambda$.

\topinsert
  \vbox{\tabskip=0pt \offinterlineskip
  \def\tablerule{\noalign{\hrule}}
  \halign to 5.5in {\strut#& \vrule#\tabskip=1em plus2em&
    \hfil#& \hfil#& $\sim$ #& #\hfil& \vrule#&
    #\hfil& \vrule#\tabskip=0pt\cr\tablerule
  &&\multispan4&& \omit\hidewidth $\lambda\sim g^3$ \hidewidth&\cr
  \tablerule
  &&    & $\phi$              && $(g^3/\lambda)T$       && $T$         &\cr
  && scalar mass
        & $m$                 && $(g^3/\sqrt\lambda)T$  && $g^{3/2}T$  &\cr
  && transverse vector mass
        & $M$                 && $(g^4/\lambda) T$      && $gT$        &\cr
  && vector Debye mass
        & $\sqrt{\ML^2-M^2}$  && $gT$                   && $gT$        &\cr
  && vector loop expansion parameter
        & $g^2T/M$            && $\lambda/g^2$          && $g$         &\cr
  && scalar loop expansion parameter
        & $\lambda T/m$       && $(\lambda/g^2)^{3/2}$  && $g^{3/2}$   &\cr
  &&
        & $(\Tc-T_0)/\Tc$     && $g^4/\lambda$          && $g$         &\cr
  && barrier in $V$ between minima
        & $V$                 && $(g^{12}/\lambda^3)T^4$  && $g^3T^4$    &\cr
  && 2-loop correction to potential
        & $(g^2T/M)V$         && $(g^{10}/\lambda^2)T^4$  && $g^4T^4$    &\cr
  \tablerule}}
  \openup-3\jot
  \noindent  {\bf Table 1.}
  Orders of magnitude of parameters in the $\phi\not=0$
  vacuum at the phase transition.  The far right column gives the
  simplified power-counting rules, which assume $\lambda\sim g^3$.
  The entry for the 2-loop correction to the potential refers to
  the dominant $\phi$-{\it dependent} corrections.
  \vskip 0.5in
\endinsert

Assuming $g^4 \ll \lambda \ll g^2$, our goal is to consistently compute
the leading correction to the ring-improved one-loop potential.
It is cumbersome to compare orders of magnitude of various corrections
when there are two coupling constants $g$ and $\lambda$.
Fortunately, the power counting can be simplified by formally taking
\eqn\lamOrder{
  \lambda\sim g^3,
}
which is at the geometric center of the range
$g^4 \ll \lambda \ll g^2$ under consideration.  This simplification
seems to always order the relative size of corrections in a way that
is consistent over this entire range of $\lambda$.  We henceforth
always assume $\lambda \sim g^3$ unless otherwise stated.
The simplified power-counting is shown in the far-right column of
Table 1.  Note that the vector loop expansion parameter is $g$
(rather than $g^2$, as it would be at zero temperature)
and that the Debye mass and the transverse vector mass are
formally the same order of magnitude.

\subsec{Multiple loops and IR disasters.}

In studies of high-temperature QCD, it is well-known that
the perturbative computation of the free energy breaks down at
three loops because of infrared divergences associated with
static ($k_0=0$), transverse gluons.\refmark\reviews\
Though the longitudinal
gluons pick up a Debye mass of order $gT$ at one loop, as discussed earlier,
the transverse gluons do not.
These IR divergences are interpreted as a sign that the transverse
gluons must have a mass of order $g^2 T$ which is not perturbatively
calculable, corresponding to a finite screening length for static,
magnetic fields in the plasma.  The uncertainty in the free energy
due to the incalculability of this effect is
order $g^6 T^4$.\foot{
   We remind the reader of a simple mnemonic for this result.
   The only scale in the
   effective three-dimensional theory, which is the source of the IR
   divergences, is $g^2 T$.  So the free energy of the three-dimensional
   theory, which will be incalculable by perturbation theory, is order
   $(g^2 T)^3$ by dimensional analysis.  The relation between the
   4-dimensional and 3-dimensional free energies is a factor of $T$,
   giving $g^6 T^4$.}

The situation is different in a spontaneously broken theory because
the gauge bosons corresponding to broken symmetries are not massless.
In pure electroweak theory, for example, the transverse W and Z bosons
are massive in the symmetry-breaking minimum at the phase transition.
One can compute to {\it any} order of perturbation theory in this
minimum without infrared singularities.  (A perturbative
expansion is still not useful, of course, unless the loop expansion
parameter is small.)
However there is still a limit to how well the phase transition can
be studied, because determining the temperature of the phase transition
requires comparing the free energy of the asymmetric $\phi\not=0$ minimum
with the symmetric $\phi=0$ one, and all the familiar problems
of high-temperature QCD arise in the symmetric minimum.  The
calculation of the free energy at $\phi=0$ breaks down at three loops
and has an uncertainty of order $g^6 T^4$; there is therefore
little point in computing the free energy of the asymmetric vacuum to
any better accuracy.  At the phase transition, the typical size of each
of the three terms in \Vcubic\ is $g^3T^4$ and the
size of $n$-loop corrections are suppressed by $(g^2 T/M)^{n-1}$,
giving $g^{n+2} T^4$.
So there is no point computing the free energy beyond three loops near
the asymmetric vacuum.
Being unadventurous, we shall limit ourselves to
computing 2-loop corrections.

%

\newsec{Warming Up With a Scalar Toy Model}

\seclab\ToySec

We now want to proceed to compute the dominant 2-loop contributions
to the potential.  To introduce the basic computational method, we
wish to start with as simple an example as possible: a
pure scalar theory of a single, real field.  Sadly, we saw in the
last section that the loop expansion cannot be trusted to study
the phase transition in this model, because the loop expansion
parameter $\lambda T/\meff(\phic)$ is order 1.  However there is
no reason we cannot compute the effective potential at larger
values of $\phi$ where $\meff(\phi)$ is larger so that the loop
expansion parameter is smaller.
This region of the potential has nothing to do
with the phase transition in this model, but it provides a simple
example of the computations and approximations that we later implement
in gauged models.  We shall take a classical Higgs potential of the
form
\eqn\ghiggs{
  \Vcl(\phi) = -{1\over2}\mmu^2\phi^2 + {1\over4!} g^2 \phi^4
}
and shall study the potential for $\phi$ of order
\eqn\PhiToy{
  \phi \sim T
}
at temperatures $T$ near the critical temperature $\Tc$.
This is supposed to be analogous to the
$\lambda\sim g^3$ gauge theory discussed in the last section insofar as
(1) there is a zero-temperature loop expansion parameter called $g^2$,
and (2) we examine the potential at $\phi\sim T$ (see Table 1).
The effective scalar
mass is then of order $gT$ and the loop expansion parameter is
order $g^2 T/ m\sim g$, both analogous to the vector mass and vector
loop expansion parameter in Table 1.  We have named the scalar
self-coupling $g^2$ in the toy model both to emphasize this
analogy and to avoid confusion with our $\lambda\sim g^3$
convention used in the gauged models.  We hope that the reader's benefit
from seeing the techniques first presented in a simple model
will outweigh any confusion inherent in the temporary change of conventions.
Orders of magnitude are summarized in Table 2.

\topinsert
  \vbox{\tabskip=0pt \offinterlineskip
  \def\tablerule{\noalign{\hrule}}
  \halign to 4.5in {\strut#& \vrule#\tabskip=1em plus2em&
    \hfil#& \hfil#& #& #\hfil& \vrule#\tabskip=0pt\cr\tablerule
  &&\multispan4 \hfil SCALAR TOY MODEL \hfil&\cr
  \tablerule
  && by assumption:   & $\phi$              & $\sim$  & $T$       &\cr
  && effective mass at $\phi$
                      & $m(\phi)$           & $\sim$  & $gT$      &\cr
  && Debye mass       &                     & $\sim$  & $gT$      &\cr
  && loop expansion parameter
                      & $g^2T/m(\phi)$      & $\sim$  & $g$       &\cr
  && 2-loop correction to potential
                      &                     & $\sim$  & $g^4T^4$    &\cr
  \tablerule}}
  \openup-3\jot
  \noindent  {\bf Table 2.}
  Orders of magnitude of parameters in the scalar toy model for
  $\phi \sim T$ and $T \sim \Tc$.
  The entry for the 2-loop correction to the potential refers to
  the dominant $\phi$-{\it dependent} corrections.
  \vskip 0.5in
\endinsert

Let us apply our earlier review of power counting to estimate the
two-loop contribution to the potential.  Ignoring quadratically divergent
loops, which will be absorbed by ring-improvement, the dominant
two-loop contribution
should be order $g^2 T^2 m^2$, where there is an explicit power of $g^2$,
an explicit power of $T$ for each loop, and then $m^2$ by dimensional analysis.
Since $m\sim gT$, the 2-loop contribution is
equivalently order $g^4 T^4$.  When computing the 2-loop potential, we shall
always drop any contributions smaller than $O(g^4 T^4)$.
In the region of $\phi$ under study, the contributions dropped are of the
same order as three- and higher-loop contributions, and so there is no
point retaining them in a two-loop calculation.

\subsec{The 1-loop result without resummation}

\subseclab\ToySecA

We shall leave the analysis of resumming rings for sections 3.3 and 3.4.
For now, we focus on the details of the unimproved loop expansion
in this model and start by carefully examining one-loop results.
We need to keep terms to order $g^4 T^4$, which requires one higher
order in the high-temperature expansion \SVone{a}\ discussed in the
introduction.  At this order the one-loop potential receives UV
infinite contributions from usual zero-temperature divergences, and
so we must confront regularization and counter-terms.
We shall regularize all our calculations using dimensional regularization
in $4-2\eps$ dimensions.
Computing counter-terms to one-loop order, one finds that the bare
potential expressed in terms of the renormalized coupling constants is
\eqn\Vbare{
   \eqalign{
     V_\bare(\phi) &=
        - {1\over2}\mmu_\bare^2\phi_\bare^2
        + {1\over4!}g_\bare^2\phi_\bare^4
   \cr
     &= - {1\over2} Z_1 \mmu^2\phi^2
        + {1\over4!} \mu^{2\eps} Z_2 g^2\phi^4 ,
   }
}
\eqna\Ztoy
$$
   \eqalignno{
      Z_1 &= 1 + {1\over32\pi^2}g^2{1\over\eps} + O(g^4) ,
   &\Ztoy a\cr
      Z_2 &= 1 + {3\over32\pi^2}g^2{1\over\eps} + O(g^4) .
   &\Ztoy b\cr
   }
$$
$\mu$ is the arbitrary renormalization scale.
We shall use \MSbar\ regularization, which, we remind the reader, corresponds
to performing minimal subtraction (MS) and then changing scales to $\mub$
by the substitution
\eqn\MSscale{
   \mu = \bar\mu\left(e^{\gammaE}\over4\pi\right)^{1/2} ,
}
where $\gammaE$ is Euler's constant.

The next term in the high-temperature expansion of the one-loop potential
is well-known
(but we shall review the derivation in section 3.4):\refmark\dolan
\eqn\VtoyOne{
   \mu^{2\eps}V^{(1)}_\naive(\phi) = \mu^{2\eps}V_\bare(\phi) + J[m(\phi)],
}
\eqn\eqJ{
   \eqalign{
      J(m) &= {1\over2} \mu^{2\eps} \sumint_K \ln(K^2+m^2)
   \cr
           &= \const + {1\over24} m^2 T^2 - {1\over12\pi} m^3 T
                     - {1\over64\pi^2} m^4 \left[\Jepslog\right]
   \cr & \qquad\qquad
                     + O(m^6/T^2) + O(\eps).
   }
}
where
\eqn\mdefn{
   m^2(\phi) = -\mmu^2 + {1\over2}g^2\mu^{2\eps}\phi^2,
   \qquad
   \cb = \ln(4\pi)-\gammaE,
}
and the integral-summation sign above is short-hand for the Euclidean
integration
\eqn\eqSumint{
   \sumint_K \rightarrow T \sum_{k_0}
      \int { \d^{3-2\eps} k \over (2\pi)^{3-2\eps} } .
}
The sum is over $k_0 = 2\pi n T$ for all integers $n$.

The divergences in $J(m)$ cancel against the counter-terms in
$V_\bare(\phi)$.
The reader should note that our definition of $J(m)$ includes
{\it both} the temperature-dependent and zero-temperature
contributions.\foot{
  One often sees only the temperature-dependent piece in the literature,
  which has an $m^4 \ln(m^2/T^2)$ term, which is not analytic in $m^2$.
  It is important to realize that in the full result, which includes the
  zero-temperature contribution, the only term not analytic in $m^2$
  is the $m^3 T$ term. }
Note also that \MSbar\ regularization does
not get rid of all factors of $\gammaE$ and $\ln(4\pi)$ as it would
at zero temperature.
The subscript $\naive$ above refers to the absence of resummation.

\subsec{The 2-loop result without resummation}

\subseclab\ToySecB

\PSfig\figh{
  2-loop contributions to the potential of the scalar toy model.
}{4.7cm}{fig7.psx}{80}

The 2-loop
diagrams are shown in \figh,
where the heavy dots represent
1-loop, zero-temperature counter-terms.  The propagators use the
mass \mdefn\
appropriate in the presence of the background field $\phi$, and we
shall usually refer to $m^2(\phi)$ simply as $m^2$.  The crosses in
the diagrams represent explicit factors of the background field
$\phi$ at vertices.

Fig.~\xfig\figh a is the most straightforward of the 2-loop diagrams, giving a
contribution to the effective potential of
\eqn\Vtoya{
   \mu^{2 \eps}V^{\rm(a)}_\naive = {1\over8} g^2 \left[ I(m) \right]^2
}
where
\eqn\defI{
   I(m) = \mu^{2\eps} \sumint_K {1 \over K^2+m^2} .
}
$I(m)$ is related to $J(m)$ of \VtoyOne\ by
$I(m) = m^{-1}\,\d J/\d m$, and so its high-temperature expansion is
\eqn\eqI{
   \eqalign{
     I(m) = {1\over12} T^2
           &- {1\over4\pi} m T
           - {1\over16\pi^2} m^2 \left[\Jepslog\right]
           + O(m^4/T^2)
   \cr&
           + \eps \ceps {1\over 12} T^2 + O(\eps m T) ,
   }
}
where we have now explicitly shown the leading term of order $\eps$.
Though it will appear in results for individual diagrams, it turns out
that
\eqn\eqceps{
   \ceps = \ln\left(\mub^2\over T^2\right) + 2\gammaE
           - 2\ln2 - 2{\zeta'(2)\over\zeta(2)}
}
is an unimportant constant because it will cancel in our final
result.  (A sketch of the derivation of $\ceps$ may be found in section 3.4.
$\zeta$ is the Riemann zeta function.)
Using the expansion \eqI\ in the contribution \Vtoya\ to the
potential, and keeping terms only up to $O(g^4 T^4)$,
\eqn\VVtoya{
   \eqalign{
      \mu^{2\eps}V^{\rm(a)}_\naive =
         \const &- {1\over48\cdot4\pi} g^2 m T^3
                 - {1\over48(4\pi)^2} g^2 m^2 T^2
                      \left[ \Jepslogc - 6 \right]
   \cr&
             + O(g^5 T^4) .
   }
}
(When we later address the resummation of rings, we shall find that
the first term, which is order $g^3 T^4$, is absorbed by
the resummation.)

Fig.~\xfig\figh b is more interesting because it is logarithmically divergent
in the three-dimensional theory (that is, when all loop frequencies
$k_0$ are set to zero).  As a result, it generates a logarithmic
dependence on the mass $m(\phi)$, unlike the one-loop contribution or
\figh a.  As we shall see when we later return to gauge theories,
it is such logarithmic terms which are solely responsible, at the order
under consideration, for modifying the VEV at the phase transition from
its one-loop value.

Turning to specifics, the contribution of \figh b is
\eqn\Vtoyb{
   \mu^{2\eps}V^{\rm(b)}_\naive = -{1\over 12} g^4 \mu^{2\eps} \phi^2 H(m) ,
}
where
\eqn\defH{
   H(m) = \mu^{4\eps} \sumint_P \sumint_Q
           { 1
             \over
             (P^2+m^2) (Q^2+m^2) [(P+Q)^2+m^2] } .
}
The high-temperature limit of $H$ has been evaluated by
Parwani,\refmark\parwani\ who finds
\eqn\eqH{
   \eqalign{
      H(m) = {1\over64\pi^2} T^2 &\left(
                 {1\over\eps} + \ceps + \ln\left(\mub^2 \over T^2\right)
                 + 2\ln\left(T^2 \over m^2\right) + 2 - \cH \right)
   \cr &
           + O(m^2) + O(\eps T^2)
   }
}
where
\eqn\eqcH{
  \cH \approx 5.3025
}
is a numerical constant which can be expressed in terms of
double definite integrals of elementary functions.
(See Ref.~\parwani\ for details.)
Substitution into \Vtoyb\ gives
\eqn\VVtoyb{
   \mu^{2 \eps}V^{\rm(b)}_\naive =
      {1\over48(4\pi)^2} g^4 \mu^{2\eps}\phi^2 T^2
           \left[
              -{1\over\eps} -\ceps - \ln\left(\mub^2 \over T^2\right)
              - 2\ln\left(T^2 \over m^2\right) - 2 + \cH \right]
       + O(g^5 T^4) .
}
Note the promised presence of $\ln[m(\phi)]$.

The two counter-term graphs figs.~\xfig\figh c and \xfig\figh d give
\eqn\Vtoyc{
      \mu^{2 \eps}V^{\rm(c)}_\naive
      = - {1\over4(4\pi)^2} g^2 \mmu^2 I(m) {1\over\eps}
      = \const
         + O(g^5 T^4) ,
}
\eqn\Vtoyd{
      \mu^{2 \eps}V^{\rm(d)}_\naive =
          {3\over8(4\pi)^2} g^4 \mu^{2\eps}\phi^2 I(m) {1\over\eps}
      = {3\over96(4\pi)^2} g^4 \mu^{2\eps}\phi^2 T^2
            \left({1\over\eps}+\ceps\right)
         + O(g^5 T^4) .
}
Combing the results of this subsection with \VtoyOne\ yields the
unimproved two-loop potential:
\eqn\VtoyTwo{
   \eqalign{
      \mu^{2\eps}V^{(2)}_\naive =
      \const &- {1\over48\cdot4\pi} g^2 m T^3
   \cr &
      + {1\over2} \biggl\{
            \left( - \mmu^2(T) + {1\over24} g^2(T) T^2 \right)
            - {1\over(4\pi)^2} \cb g^2 \mmu^2
      \cr & \qquad\qquad
            - {1\over24(4\pi)^2} g^4 T^2 \left[
                2 \ln\left(T^2\over m^2\right)
                - 1 - \cb - \cH
              \right]
            \biggr\} \phi^2
   \cr &
       - {1\over12\pi} m^3 T
       + {1\over4!} \left[ g^2(T) + {3\over(4\pi)^2} \cb g^4
         \right] \phi^4
       + O(g^5 T^4)
   }
}
where
\eqn\mdefb{
   m^2 = -\mmu^2 + {1\over2}g^2\phi^2 ,
}
\eqn\runtoyg{
   g^2(T) = g^2 - {3\over32\pi^2} g^4 \ln\left(\mub^2\over T^2\right)
                + \cdots ,
}
\eqn\runtoymmu{
   \mmu^2(T) = \mmu^2 \left[ 1
         - {1\over32\pi^2} g^2 \ln\left(\mub^2\over T^2\right)
         + \cdots
       \right] .
}

As it should be, the result is invariant under the renormalization group to
the order we have computed.
We have chosen to write the answer in terms of the running couplings\foot{
  These couplings are run with the usual zero-temperature renormalization
  group and do not represent the use of any sort of temperature-dependent
  renormalization group equations. }
$g^2(T)$ and $\mmu^2(T)$.  (At this order we need not worry about
the anomalous scaling of $\phi$.)
The physical scales in this problem are $m$ and $T$, and so, when
evaluating \VtoyTwo\ in practice, we should choose the renormalization
scale $\mub$ to avoid producing large logarithmic enhancements,
$\ln(\mub/T)$ or $\ln(\mub/m)$, of yet higher-order corrections.
Fortunately, since $m \sim gT$, $m$ and $T$ are not drastically
different scales.
The difference between $g^2(T)$ and $g^2(m)$ is of order
$g^4 \ln g$ and is indeed small, and so $\mub \sim T$ and
$\mub \sim m$ are both adequate choices.  We chose to write
\VtoyTwo\ in terms of $g^2(T)$ instead of $g^2(m)$, but this
choice is merely convention and lacks physical significance.

\subsec{Resummation: method I}

\subseclab\ToySecC

How do I resum thee?  Let me count the ways.

We now want to implement the resummation of the dominant parts of
ring diagrams by replacing $m^2(\phi)$ in our propagators by
$\meff^2(\phi)$ as defined in \Smeff\ to include the thermal
contribution to the mass.  Parwani\refmark\parwani
(who has computed the sub-leading
correction to the scalar mass at high temperatures) uses one
method for doing this systematically and consistently.
Rewrite the bare potential \Vbare\ as
\eqn\VbareR{
   V_\bare(\phi) =
      {1\over2} Z_1 \left( -\mmu^2 + {1\over24} g^2 T^2 \right) \phi^2
      + {1\over4!} \mu^{2\eps} Z_2 g^2 \phi^4
      - {1\over48} Z_1 g^2 T^2 \phi^2 ,
}
where the thermal mass term has simply been added in and subtracted out
again so that nothing is changed.  But now interpret the first term
as part of the {\it unperturbed} Lagrangian ${\cal L}_0$ and treat the
last term as a perturbation.
(We shall loosely refer to the last term as the thermal ``counter-term,''
but we have not in fact changed the renormalization prescription from
the usual zero-temperature one.)
Nothing has changed if all orders of
perturbation theory are summed.  However, order by order, the $g^2 T^2$
pieces of quadratically-divergent sub-loops now cancel against
diagrams involving the thermal counter-term; the new perturbative
expansion is controlled by the convergent-loop expansion parameter
$g^2 T/m$.

\PSfig\figi{
  2-loop contribution involving the thermal counter-term.
}{18cm}{fig8.psx}{25}

In the case at hand, this resummation replaces $m(\phi)$ by
\eqn\meffdef{
   \meff(\phi) = - \mmu^2 + {1\over2}g^2\mu^{2\eps}\phi^2
                 + {1\over24}g^2 T^2
}
in our previous calculations of contributions to the two-loop potential\foot{
  This substitution may be made directly in the results of sections \ToySecA\
  and \ToySecB\ except for the final formula \VtoyTwo\ because the expansion
  \mdefn\ of $m^2(\phi)$ was used when putting the result in its final form.
}
and introduces the
new diagram of \figi.
The box represents the thermal
counter-term---the interaction generated by the last term in \VbareR.
Fig.~\xfig\figi\ gives
\eqn\Vtoyx{
   \eqalign{
      \mu^{2 \eps}V^{\rm(c.t.)}
      &= - {1\over48} g^2 T^2 I(\meff)
   \cr
      &= \const + {1\over48\cdot4\pi} g^2 \meff T^3
                + {1\over48(4\pi)^2} g^2 \meff^2 T^2 \left[\Jepslog\right]
   \cr & \qquad
                + O(g^5 T^4) .
   }
}
Replacing $m\rightarrow\meff$ in the original 2-loop contributions
\VVtoya, \VVtoyb--\Vtoyd\ and combining with \Vtoyx\ above now gives
\eqn\VtoyTwoR{
   \eqalign{
      V^{(2)} =
      \const
   &
      + {1\over2} \biggl\{
            \left( - \mmu^2(T) + {1\over24} g^2(T) T^2 \right)
            - {1\over(4\pi)^2} \cb g^2 \mmu^2
   \cr & \qquad\qquad
            - {1\over24(4\pi)^2} g^4 T^2 \left[
                2 \ln\left(T^2\over \meff^2\right)
                - 1 - \cb - \cH
              \right]
            \biggr\} \phi^2
   \cr &
       - {1\over12\pi} \meff^3 T
       + {1\over4!} \left[ g^2(T) + {3\over(4\pi)^2} \cb g^4
         \right] \phi^4
       + O(g^5 T^4)
   }
}
Compare this to the unimproved result \VtoyTwo.
As promised earlier, the $O(g^2mT^3)$ term of the figure-eight diagram
\VVtoya\ has disappeared, canceled by the thermal counter-term diagram
\Vtoyx.  The original $O(g^2mT^3)$ contribution arose from taking the
sub-leading $O(mT)$ contribution of one loop integral times the
leading $O(T^2)$ piece of the other, and the latter is precisely
what resummation is intended to absorb.

\subsec{Resummation: method II}

\subseclab\ToySecD

The implementation of resummation above is a little less natural for
gauged theories.  The scalar ring-diagram in \figa\ is independent
of momentum, whereas the quadratically-divergent diagrams of \figf\
for gauged theories are not.  Moreover, the polarization dependence of
these $g^2T^2$ contributions to the vector self-energy $\Pi_{\mu\nu}(K)$
{\it also} depends on momentum\refmark\reviews\
and is simple only in limits such as $K\rightarrow 0$,
where only the contribution to $\Pi_{00}(0)$ is non-zero.\foot{
  We emphasize that we are working in Euclidean space, and so by
  $\Pi(0)$ we always mean $\Pi(k_0=0, \vec k\rightarrow 0)$ since $k_0$
  is discrete.  If one analytically continues $k_0$ to real-time and takes
  instead the limit $\Pi(k_0 \rightarrow 0, \vec k = 0)$, the limit
  is completely different, giving the plasma mass for propagating
  waves $\vec A$ rather than the Debye mass for static
  electric potentials $A_0$. }
Having computed the leading contribution
to the self-energy $\Pi(K)$, should one resum the vector propagator
$G_0(K)$ as $1/[G_0^{-1}(K) + \Pi(K)]$ or $1/[G_0^{-1}(K) + \Pi(0)]$ or
something else?
The answer is that it doesn't matter.
Resummation only affects perturbative expansions when $\Pi$ cannot
be treated as a perturbation to the inverse propagator $G_0^{-1}$.
This happens only when $K^2 \ll T^2$,
in which case $\Pi(K) \approx \Pi(0)$.

\PSfig\figj{
  Cancellations of resummation and thermal counter-terms.
}{2cm}{fig9.psx}{90}

To see more explicitly that resummation is irrelevant when
$\Pi \ll G_0^{-1}$, consider $K^2 \sim T^2$.
Then $\Pi(K) \sim g^2 T^2$ and $G_0^{-1}(K) \sim T^2$.
The resummed propagator for $K^2 \sim T^2$ can then be expanded
perturbatively in powers of $\Pi G_0 \sim g^2$.  For example,
the resummed propagator of the previous section may be expanded
as\foot{
  Since $m^2 \sim g^2T^2$, we could expand the powers of $m$ as well,
  but we have not bothered to do so. }
\eqn\expandG{
   {1\over K^2 + m^2 + a g^2 T^2}
   =
   {1\over K^2 + m^2} - {a g^2 T^2 \over (K^2 + m^2)^2}
        + {(a g^2 T^2)^2 \over (K^2 + m^2)^3} - \cdots
}
The $n$-th term is order $g^{2n}$.
Every term, except the first, gives contributions (inside any diagram) that
cancel {\it order-by-order}
against insertions of the thermal counter-term, as in \figj.
So resummation for $K^2 \sim T^2$ has no effect on the perturbative
expansion of the final result.  (We shall see this even more
explicitly in a moment by recomputing the 2-loop potential in
our scalar toy model using a different resummation prescription.)

The circumstance where resummation is relevant is the infrared
region $K^2 \lsim (gT)^2$ where $\Pi \sim G_0^{-1}$.
Then all terms in \expandG\ are the same order.
Any resummation prescription which approximates $\Pi(0)$ in the
infrared region will do.
In general, any resummation
$1/[G_0^{-1}(K) + {\cal P}(K)]$ will work if
(1) ${\cal P}(K) \approx \Pi(0)$ when $K^2 \lsim (gT)^2$
and (2) ${\cal P}(K) \lsim g^2T^2$ for general $K$.

Since $K^2 \lsim (gT)^2$ implies $k_0 = 0$ in Euclidean space,
the resummation prescription for $k_0 \not= 0$ modes is irrelevant
(as long as condition (2) is met).
Since resummation is an infra-red phenomena, we find it convenient
to adopt the following point of view, in the spirit of decoupling
and the renormalization group.  When computing the effective potential,
first integrate out all of the heavy $k_0 \not=0$ modes (ignoring the issue
of resummation) to obtain an effective,
three-dimensional theory of the $k_0 = 0$ modes.
This will generate the thermal mass terms as well as other interactions
induced by the heavy modes, which we compute to the desired order
in perturbation theory.  Only then do we finally integrate the
$k_0 = 0$ modes after deciding on a sensible partition of the
effective three-dimensional Lagrangian into an unperturbed piece
${\cal L}_0$, containing the thermal mass terms, and a
perturbative piece.  In the language of the previous paragraphs, this
corresponds to choosing a resummed propagator $1/(G_0^{-1} + {\cal P})$
with
\eqn\resumpre{
   {\cal P}(K) = \cases{
                    \bar\Pi(0) , & $k_0 = 0$ ,     \cr
                    0          , & $k_0 \not= 0$ , \cr
                 }
}
where $\bar\Pi(0)$ is the dominant, $O(g^2 T^2)$ term of $\Pi(0)$.
(One could alternatively replace $\bar\Pi(0)$ by the full $\Pi(0)$
calculated to some order in perturbation theory,
but perturbative changes to ${\cal P}$ have no effect on the
perturbation expansion.)

Though resumming only the $k_0=0$ modes sounds a little more cumbersome
than resumming all the modes, we find it algebraically simpler in
the gauge theory case because $1/[G_0^{-1}(K) + \Pi(K)]$ and
$1/[G_0^{-1}(K) + \Pi(0)]$ turn out to have more complicated polarization
dependence for $k_0 \not = 0$ than for $k_0 = 0$.  We also find it
conceptually simpler to apply resummation only to the modes
which require it.

As a paradigm for splitting calculations into heavy $k_0 \not= 0$ modes
and light $k_0 = 0$ modes,
and to emphasize the fact that masses (or self-energies)
may be treated perturbatively for heavy modes but not for light ones,
we shall briefly
sketch the derivation of the high-temperature expansion \eqI\ of
$I(m)$ from its definition \defI.
\eqn\deriveI{
  \eqalign{
     I(m)
  &=
     \mu^{2\eps} T \intk {1\over k^2+m^2}
     + 2 \mu^{2\eps} T \nsum\intk {1\over (2\pi nT)^2 + k^2 + m^2}
  \cr &=
     \mu^{2\eps} T \intk {1\over k^2+m^2}
     + 2 \mu^{2\eps} T \lsum\nsum\intk
        {(-)^l m^{2l}\over [(2\pi n T)^2 + k^2]^{l+1}}
  \cr &=
     { \Gamma\left(-{1\over2}+\eps\right) \over (4\pi)^{3\over2} }
       \left( {4\pi\mu^2\over m^2}\right)^\eps m T
  \cr & \qquad\qquad\qquad
     + {T^2\over (4\pi)^{1\over2}} \left( {\mu^2\over \pi T^2}\right)^\eps
       \lsum { \Gamma\left(l-{1\over2}+\eps\right) \over\Gamma(l+1) }
       (-)^l \left({m\over 2\pi T}\right)^{2l} \zeta(2l-1+2\eps)
  \cr &=
     {1\over12} T^2
     - {1\over4\pi} m T
     - {1\over16\pi^2} m^2 \left[\Jepslog\right]
  \cr &\qquad\qquad\qquad
     + T^2 \sum_{l=2}^\infty \left(-m^2\over4\pi^2T^2\right)^l
        {(2l-3)!!\over(2l)!!}
        \zeta(2l-1)
     + O(\eps) ,
  }
}
where $\zeta$ is the Riemann zeta function.  In the first line
we split the calculation into $k_0=0$ and $k_0\not=0$ modes; in
the second line we expand the heavy modes in powers of $m^2$;
in the third we first evaluate the momentum integrals and then sum over $n$;
and in the last line we take the $\eps \rightarrow 0$ limit.
(This derivation may be used to obtain the constant $\ceps$ of
\eqceps.)
The point of this
example lies not in the mathematical details but in the fact that,
in the sum over heavy modes,
we are able to treat $m^2$ as a perturbation; the
expansion of the integrand in $m^2$ for these modes is equivalent
to the high-temperature expansion in $m^2/T^2$.
The $k_0 = 0$ term, on the other hand, produces the
only term not analytic in $m^2$---the $mT$ term at $O(\eps^0)$.
(One may now also obtain
the high-temperature expansion \eqJ\ of $J(m)$ by using
$I(m) = m^{-1} \d J/\d m$.)

We now demonstrate how the new method of resummation works in our scalar
toy model.  For the scalar case, the new method will seem more convoluted
than the previous one, but we find it simpler to use when we compute in
gauge theories.

\PSfig\figk{
  Resummation of \figh b.  Double lines are resummed, ``heavy'' denotes
  $k_0\not=0$ modes and ``zero'' denotes $k_0=0$ modes.
}{2.5cm}{fig10.psx}{90}

The resummation of the setting-sun diagram of \figh b
is shown in \figk.
The $k_0 \not= 0$ lines are marked
``heavy'' and the $k_0 = 0$ lines marked ``zero''.  The double lines
represent resummed propagators.
This expression may be simplified by realizing that the second
term is order $g^5 T^4$ and hence ignorable; this term is
of the form (leaving out $\eps$'s to avoid clutter):
\eqn\bloop{
  \eqalign{
  &
     {1\over4} g^4\phi^2 T \intpthree {1\over(p^2 + \meff^2)}
       T \sum_{k_0\not=0} \intkthree {1\over (K^2 + m^2) [(P+K)^2 + m^2]}
  \cr &=
     {1\over4} g^4\phi^2 T \intpthree {1\over(p^2 + \meff^2)}
       \bigl[ O(\ln(T/\mub)) + O(p^2/T^2, m^2/T^2)
  \cr & \qquad\qquad\qquad\qquad\qquad
          + O(p^4/T^4, m^4/T^4) + \cdots \bigr]
  \cr &=
     {1\over4} g^4\phi^2 T \left[ O(m\ln(T/\mub)) + O(m^3/T^2)
          + O(m^5/T^3) + \cdots \right]
  \cr &=
     O(g^5 T^4 \ln(T/\mub)) + O(g^7 T^4) + O(g^9 T^4) + \cdots .
  }
}
Note that we have implicitly used dimensional regularization to make
sense of UV divergent integrals such as $\int \d^3p\,p^n/(p^2+m^2)$
and estimate their order as $m^{n+1}$.\foot{
  The reader may be concerned by this.
  It is possible, but cumbersome, to reorganize the 3-dimensional
  ($k_0 = 0$)
  integrals that we shall encounter in implementing resummation at
  2-loops so that they are UV convergent.  For instance, the difference
  between
  the resummed and unresummed results takes the form
  $\int \d^3p\,p^n[(p^2+\meff^2)^{-1}-(p^2+m^2)^{-1}]$, which is
  more UV convergent.  Also, we are only interested in the $\phi$ dependence
  of the potential, and so may instead evaluate $\d/\d\phi$ of the
  potential, which makes the UV behavior even more convergent.
  With enough care, one may draw the same conclusions about which
  terms are negligible from UV convergent integrals.
}

\PSfig\figl{
  Relation between a heavy contribution and the unresummed graph.
}{2.5cm}{fig11.psx}{90}

\PSfig\figm{
  Rewriting of \figk.
}{4cm}{fig12.psx}{90}

We may compute heavy-loop diagrams, like the first term in \figk,
by relating them
to our results from un-resummed perturbation theory.
Fig.~\xfig\figl\ shows such a relation.  None of the lines are resummed,
and unlabeled lines represent the sum over {\it all} values of
$k_0$.  The first term is the previous, unresummed result of
\VVtoyb, the second term is $O(g^5 T^4)$ as discussed above,
and the last term is a 2-loop diagram in three dimensions.
Combining with \figk, resummation of the setting-sun diagram
is therefore implemented by \figm.
We now need the difference
of two three-dimensional graphs, the last two terms of \figm,
which differ only by the mass used in the propagators.
One may either compute the three-dimensional integrals directly
or note that the the mass dependence in the four-dimensional result
\VVtoyb\ can only come from the $k_0 = 0$ term at this order.
So the three-dimensional graphs are each
\eqn\VVtoybThree{
   {1\over48(4\pi)^2} g^4\phi^2 T^2
      \left[ \const + 2\ln\left(T^2\over m^2\right) \right] ,
}
with $m$ replaced by $\meff$ in the third term of \figm.
The total effect of the sum of terms in \figm\ is therefore to simply
replace $m$ by $\meff$ in the unresummed, four-dimensional result
\VVtoyb:
\eqn\VVtoybR{
   \mu^{2 \eps}V^{\rm(b)} =
      {1\over48(4\pi)^2} g^4 \mu^{2\eps}\phi^2 T^2
           \left[
              {1\over\eps} + \ceps - \ln\left(\mub^2 \over T^2\right)
              + 2\ln\left(T^2 \over \meff^2\right) + 2 - \cH \right]
       + O(g^5 T^4) .
}

\PSfig\fign{
  Resummation of \figh a.
}{4.8cm}{fig13.psx}{90}

The resummation of the figure-eight diagram is shown in \fign.
The result is easily evaluated by referring to the review
\deriveI\ of the expansion of $I(m)$.  The three-dimensional
piece of $I(m)$ is the $mT$ term,
\eqn\Ithree{
   I_3(m) = -{1\over4\pi} mT ,
}
and the contribution due to heavy modes is everything else,
\eqn\Iheavy{
  \eqalign{
     I_{\rm h}(m) =
  &
           {1\over12} T^2 (1+\eps\ceps)
           - {1\over64\pi^2} m^2 \left[\Jepslog\right]
  \cr &
           + O(m^4/T^2) + O(\eps m^2).
  }
}
The resummation of the figure-eight is then
\eqn\VVtoyaR{
   \eqalign{
      \mu^{2\eps}V^{\rm(a)} =
         \const &- {1\over48\cdot4\pi} g^2 \meff T^3
                 + {1\over8(4\pi)^2} g^2 \meff^2 T^2
   \cr&
                 - {1\over48(4\pi)^2} g^2 m^2 T^2
                     \left[ \Jepslogc \right]
             + O(g^5 T^4) .
   }
}
Note that this is not simply the substitution of $m\rightarrow\meff$
into the unresummed result.

Resummation of the counter-term diagrams
of \figh c and \figh d do not change the unresummed results
\Vtoyc\ and \Vtoyd\ through $O(g^4 T^4)$.  The thermal-counter term
of \figi\ now only applies to the $k_0=0$ modes, giving
\eqn\Vtoyx{
      \mu^{2 \eps}V^{\rm(c.t.)}
      = {1\over48\cdot4\pi} g^2 \meff T^3 .
}
For the one-loop result of \VtoyOne\ and \eqJ, resummation affects only
the $m^3 T$ term arising from the $k_0=0$ modes:
\eqn\VtoyOneR{
   \mu^{2\eps}V^{(1)}(\phi)
          = \mu^{2\eps}V_\bare(\phi) + J_{\rm R}[m(\phi)],
}
\eqn\eqJR{
   \eqalign{
      J_{\rm R}(m) = \const &+ {1\over24} m^2 T^2 - {1\over12\pi} \meff^3 T
                     - {1\over64\pi^2} m^4 \left[\Jepslog\right]
   \cr
                    &+ O(m^6/T^2) .
   }
}
The sum of all these contributions reproduces the previous resummed,
two-loop result \VtoyTwoR, giving an explicit example that the
exact details of the resummation prescription are unimportant.

\newsec{The Abelian Higgs Model}

We now turn to the simplest example of a spontaneously-broken gauge
theory: the Abelian Higgs model, defined by the Lagrangian
\eqn\abelL{
  {\cal L} = -{1\over4} F^2 + |D\Phi|^2 - V(|\Phi|^2) ,
  \qquad
  V(|\Phi|^2) = -\mmu^2 |\Phi|^2 + {1\over3!}\lambda |\Phi|^4 ,
}
where $\Phi$ is a complex field and
$D_\mu\Phi = (\partial_\mu - ieA_\mu)\Phi$.  We shall typically
express the potential in terms of $\phi = \sqrt2 \Phi$ so that
it takes the canonical form \higgsCl.  We now return to the original
power-counting rules of Table 1 and assume $\lambda\sim e^3$.

\PSfig\figo{
  Mixing between the $k_\mu$ polarization of $A_\mu$ and the unphysical
  Higgs boson in a background field $\phi$.
}{3.5cm}{fig14.psx}{60}

Now consider the masses of particles in a background field $\phi$.
In most gauges, such a background field induces mixing between the
scalar and the unphysical ($k_\mu)$ polarization of the vector, as
shown in \figo.
When studying effective potentials $V(\phi)$,
mixing arises even in (consistently defined) $R_\xi$ gauges because
there is no single,
$\phi$-independent gauge-fixing condition that will eliminate
mixing for all values of the background
field $\phi$.\refmark{\dolan,\dolanb}
However, there exists a gauge choice---Landau gauge---where the mixing
effectively vanishes, because in this gauge the unphysical vector
polarization does not propagate.
In order to avoid having to diagonalize propagators, which increases
the number and complexity of Feynman diagrams, we shall restrict
ourselves to Landau gauge.  Our gauge-fixing condition is therefore
\eqn\Lgf{
  {\cal L}_{\rm g.f.} = - {1\over2\xi} (\partial \cdot A)^2
       - \bar \eta \partial^2 \eta ,
  \qquad
  \xi \rightarrow 0 ,
}
where the ghost $\eta$ is completely decoupled in this Abelian model.
The Euclidean vector propagator in Landau gauge is
\eqn\Vprop{
   G_{\mu\nu}(K) = {\delta_{\mu\nu} - K_\mu K_\nu / K^2 \over K^2 + M^2 }
}
In the best of worlds it would be nice to compute
results in a variety of gauges and check that physical quantities are
indeed gauge invariant, but we have not had the perseverance
to do so.

In Landau gauge,
the masses of fluctuations in the background $\phi$ are classically
\eqn\amasses
{
  \matrix{
     M^2(\phi) = e^2\phi^2 \hfill
          & \quad ({\rm vector}) \hfill \cr
     \vphantom{\biggl[}
     m_1^2(\phi) = -\mmu^2+{1\over2}\lambda\phi^2 \hfill
          & \quad ({\rm physical~Higgs}) \hfill \cr
     m_2^2(\phi) = -\mmu^2+{1\over6}\lambda\phi^2 \hfill
          & \quad ({\rm unphysical~Goldstone~boson}) \hfill \cr
   }
}
We can now easily construct the (un-resummed) one-loop potential:
\eqn\AVone{
   \mu^{2\eps} V^{(1)}_\naive = \mu^{2\eps} V_\bare(\phi)
          + (3-2\eps) J[M(\phi)] + J[m_1(\phi)] + J[m_2(\phi)]
          + \const
}
(where the extra constant arises from the decoupled ghost contribution).
As before, the singularities in the expansion \eqJ\ of $J(m)$ cancel
against those in the bare potential $V_\bare$.
The explicit counter-terms are given in Appendix \apndxResults.

\subsec{2-loop results without resummation}

\PSfig\figp{
  2-loop diagrams in the Abelian Higgs model.
}{8.5cm}{fig15.psx}{90}

The 2-loop diagrams are shown in \figp,
where we have ignored
purely scalar diagrams since they are suppressed by $\lambda \sim e^3$
and are lower-order than $e^4 T^4$.
Ignoring resummation for now, we shall discuss how to compute the
diagrams.
Fig.~\xfig\figp a
contributes
\eqn\AVa{
  \mu^{2\eps} V^{\rm(a)}_\naive =
  - {1\over2} e^2 \mu^{4\eps} \sumint_P \sumint_Q {
     4Q^2 - 4(P\cdot Q)^2/P^2
     \over
     (P^2+M^2) (Q^2 + m_1^2) [(P+Q)^2+m_2^2] .
  }
}
This may be rewritten in terms of integrals similar to $I(m)$ and $H(m)$
of the last section by repeatedly expanding factors in the numerator
as sums of denominators, \eg
\eqn\egExpand{
  2(P\cdot Q)
  = [(P+Q)^2+m_2^2] - (P^2+M^2) - (Q^2+m_1^2) + (M^2+m_1^2-m_2^2),
}
and by noting that
\eqn\PartFrac{
  {1\over P^2(P^2+M^2)} = {1\over M^2 P^2} - {1\over M^2 (P^2+M^2)} .
}
In this way, \AVa\ may be rewritten as
\eqn\AVVa{
  \eqalign{
     \mu^{2\eps} V^{\rm(a)}_\naive
     = - {1\over2} e^2 \biggl\{ &
        I(M) [I(m_1)+I(m_2)] - I(m_1)I(m_2)
  \cr&
        + (M^2-2m_1^2-2m_2^2)\bar H(m_1,m_2,M)
  \cr&
        + {(m_1^2-m_2^2)\over M^2} [I(M)-I(0)] [I(m_1)-I(m_2)]
  \cr&
        + {(m_1^2-m_2^2)^2\over M^2} [\bar H(m_1,m_2,M) - \bar H(m_1,m_2,0)]
     \biggr\} ,
  }
}
where $I(m)$ is \defI\ as before and
\eqn\defHbar{
   \bar H(m_1,m_2,m_3)
   = \mu^{4\eps} \sumint_P \sumint_Q
           { 1
             \over
             (P^2+m_1^2) (Q^2+m_2^2) [(P+Q)^2+m_3^2] } .
}
$\bar H$ is simply a generalization of the $H(m)$ of the last section
to the case of unequal masses.

We now need a high-temperature expansion of $\bar H$.  It is easy to
relate $\bar H$ to $H(m)$ by considering the difference
\eqn\Hdiff{
   \eqalign{
     \bar H(m_1,m_2,m_3) - H(m)
     = \mu^{4\eps} \sumint_P \sumint_Q &\biggl\{
           { 1 \over (P^2+m_1^2) (Q^2+m_2^2) [(P+Q)^2+m_3^2] }
   \cr &\qquad
         - { 1 \over (P^2+m^2) (Q^2+m^2) [(P+Q)^2+m^2] }
       \biggr\}
   }
}
for any choice of masses small compared to $T$.  The difference is more
UV convergent than either alone, and the contributions to the difference
involving heavy ($p_0 \not= 0$ or $q_0 \not= 0$) modes are suppressed
by the small masses.
The dominant contribution to the difference comes from the
three-dimensional piece ($p_0=q_0=0$) and is order $T^2$.
The simplest way to evaluate it
is to switch to configuration space,\foot{
   We thank Lowell Brown for this observation.
}
and set $\eps$ to zero since the difference \Hdiff\ is UV convergent
in three dimensions.  The three-dimensional propagator is simply
$e^{-mr}/4\pi r$, and we obtain
\eqn\Hdiffb{
   \eqalign{
      \bar H(m_1,m_2,m_3) &- H(m)
   \cr
      &= {T^2\over(4\pi)^2} \int\nolimits_0^\infty
        {\d r\over r} \left\{ \exp[-(m_1+m_2+m_3)r] - \exp(-3mr) \right\}
        + O(m^2)
   \cr
      &= {T^2\over(4\pi)^2} \left\{ -\ln(m_1+m_2+m_3) + \ln(3m) \right\}
        + O(m^2) .
   }
}
Adding this to the result \eqH\ for $H(m)$ gives
\eqn\eqHbar{
   \eqalign{
      \bar H(m_1,m_2,m_3) &= H\left(m_1+m_2+m_3\over3\right)
   \cr
      &= {1\over64\pi^2} T^2 \left(
             {1\over\eps} + \ceps + \ln\left(\mub^2 \over T^2\right)
             + 4\ln\left(3T \over m_1+m_2+m_3\right)
             + 2 - \cH \right)
   \cr & \qquad\qquad
           + O(m^2) + O(\eps T^2)
   }
}
\AVVa\ may now easily be expanded:
\eqn\AVVaX{
  \eqalign{
     \mu^{2\eps} V^{\rm(a)}_\naive =
     {e^2 M T^3 \over 12(4\pi)}
  &
     - {e^2 M^2 T^2 \over 24(4\pi)^2} \left[
          {1\over\eps} + \ceps + \ln\left(\mub^2\over T^2\right)
          + 12\ln\left(3T\over M\right)
          + 6 + 4\cb - 3\cH
       \right]
  \cr&
     + O(e^{9/2} T^4) .
  }
}
The $O(e^{9/2} T^4)$ comes from terms of order $e^2 m M T$.
Though one could keep track of such terms in a 2-loop calculation,
and so improve the error to $O(e^5 T^4)$, we have not bothered
to do so.  We shall focus only on the {\it leading} correction to
the one-loop potential for $e^4 \ll \lambda \ll e^2$.

The other graphs may be computed in similar fashion, and the results
are given in Appendix \apndxResults.

\subsec{Resummation: effective masses}
\subseclab\DoubleDaisySec

We need to resum the propagators to include the dominant thermal
corrections to the masses.
We shall use the prescription of section \ToySecD\ (method II) of only
resumming the static $k_0=0$ modes.
In the Abelian Higgs model, the
longitudinal ($A_0$) mass at $(p_0=0, \vec p\rightarrow 0)$
becomes
\eqn\Aml{
   \ML^2 = e^2\phi^2 + {1\over3}e^2 T^2
}
while the transverse mass remains the same, which we shall continue to
denote by $M$:
\eqn\Amt{
   M^2 = e^2\phi^2 .
}
The leading contribution to the scalar thermal mass changes the
$m_i^2$ of \amasses\ to
\eqn\Amscal{
   \tilde m_i^2(\phi) = m_i^2(\phi)
       + \left( {1\over2}\lambda + {1\over6}\lambda + 3e^2\right)
                                                       {T^2\over12} .
}
The resummation of the 1-loop potential \AVone\ gives
\eqn\AVoneR{
  \eqalign{
     V^{(1)}
     = {1\over2} & \left[ -\mmu^2
         + \left({2\over3}\lambda+3e^2\right) {T^2\over12} \right] \phi^2
     - {1\over12\pi} (2M^3 + \ML^3) T
     + {1\over4!}\lambda\phi^4
  \cr&
     - {3 M^4\over 64\pi^2} \left[
         \ln\left(\mub^2\over T^2\right) - {2\over3} - 2\cb \right]
     + O(e^{9/2} T^4)
  }
}

\PSfig\figq{
  A generic multi-loop diagram.
}{18cm}{fig16.psx}{60}

We can now proceed as we did in section \ToySec.  Before doing so, we
should dwell a little more on the convergence of the loop expansion.
(The following discussion is important for providing a well-behaved
loop expansion but will turn out to have little practical importance
for the actual computation of the 2-loop potential to $O(e^4 T^4)$.)
Consider the multi-loop diagram of \figq.
The cost of adding loop
A to the diagram is $e^4 \phi^2 T/M^3$; the $e^4 \phi^2$ comes from
the vertices, the $T$ because it's a one-loop integral
that's not quadratically divergent,
and the $1/M^3$ to make a dimensionless quantity.  More specifically,
the $1/M^3$ arises because adding the loop added three propagators
to the diagram (two vector and one scalar) and a loop integral $\d^3 k$
which is dominated by its infrared behavior.
Both the new loop A and the loop it is attached to are dominated by
momenta of order $M$ rather than $m$, and so each propagator gives
$1/M^2$ and the new $\d^3 k$ gives $M^3$.  The total loop cost of
$e^4 \phi^2 T / M^3$ is the same order as $e^2 T/M \sim e$
(see Table 1), which is the
vector loop expansion parameter that we identified in the introduction.

Ignoring the dominant $e^2 T^2$ piece that is absorbed by resummation,
the cost of loop B is $e^2 T m / M^2 \sim e^{3/2}$
since the loops are dominated by momenta of order $m$ for the scalar and
$M$ for the vector.
Similarly, loop C cost $\lambda T / m \sim e^{3/2}$ after the dominant
$\lambda T^2$ piece is absorbed by resummation.  Adding line D cost
$e^2 T/M \sim e$.
Adding line E, however, cost $e^2 T M / m^2 \sim 1$.  So it would seem
that the loop expansion parameter is not order $e$ but instead order $1$.
The problem arises because of the difference of scales between $m$
and $M$.  When we defined the effective scalar masses $\tilde m$
in \Amscal, we only included the dominant $O(e^2 T^2)$ or
$O(\lambda T^2)$ contributions of vector or scalar loops to the
self-energy.  However, the sub-leading $O(e^2 MT)$ vector-loop
contribution is the same size as $\tilde m^2$ itself since
$\tilde m^2 \sim e^3 T^2$; it is not a perturbation and so must also
be included in the resummation.  (Yet higher-order terms {\it are}
perturbations and so do not need to be included.)
The new procedure for resummation may be viewed as follows:
first integrate out all of the heavy modes to obtain an effective
theory for momenta $k_0 = 0$ and $k \ll T$, {\it next} integrate
out the vectors (and scalars with $k \sim M$) to obtain an effective
theory for momenta $k_0=0$ and
$k \ll M$, and only then finally integrate out scalar loops
controlled by $k \sim \tilde m \ll M$.  This procedure is equivalent
to modifying our resummation prescription by including the
$O(e^2MT)$ corrections to the one-loop scalar mass \Amscal:
\eqn\goodm{
  \eqalign{
    \tilde m_i^2(\phi) \rightarrow m_i^2(\phi)
  &
       + \left( {1\over2}\lambda + {1\over6}\lambda + 3e^2\right)
                                                       {T^2\over12}
  \cr &
       - {e^2\over4\pi} (2M+\ML)T
       - {e^4 \phi^2 \over4\pi} \left({2\over M}+{1\over\ML}\right)T .
  }
}
This result may be computed directly from one-loop diagrams or, more simply,
from the second derivative through $O(e^3T^2)$ of the resummed
one-loop potential \AVoneR.
Using the resummation \goodm, the cost of adding a new loop will now always
be $\lsim e$.

\PSfig\figr{
  Generic once-iterated ring diagram for the Higgs-loop contribution
  to the effective potential.
}{15cm}{fig17.psx}{60}

Diagrammatically,
this resummation corresponds to the dominant pieces of
a subset of ``once-iterated'' ring diagrams, as shown in \figr\
for the improved scalar contribution to the one-loop
potential.\foot{
  The discussion of such diagrams appears in a related context in
  Ref.~\arnold.
}
The smallest loops of the diagram are hard, with momenta of
order $T$, the next smallest are vector loops with momenta of order
$M \sim eT$, and the large Higgs loop is softer yet, with momenta
of order $m \sim e^{3/2}T$.  Because of the hierarchy of scales,
it is a good approximation at each level of \figr\ to approximate
resummed propagators $1/[p^2+\Pi(p)]$ by $1/[p^2+\Pi(0)]$.

Though the foregoing discussion was necessary to establish an adequate
procedure for obtaining a controlled loop expansion, the details
of resumming the scalar masses turn out not to be relevant at the
order under consideration.
Though the results of some graphs have potentially significant
terms of the form $e^2mT^3$,
which would be affected by the details of how $m$ is replaced
by $\tilde m$, all such terms cancel in the final potential.
The $m^2T^2$ term of the one-loop potential is not modified to
$\tilde m^2 T^2$ if we only resum $k_0=0$ modes
(method II of section \ToySecD).\foot{
  If all modes are resummed (method I), the $m^2 T^2$ term is modified
  only by the addition of a constant, which we would ignore.
}
All other terms
involving $m$ are lower order than $O(e^4 T^4)$.
[We would need to take care to use the correct resummation \goodm\ for
$\tilde m$ if we were keeping track of $O(e^{9/2}T^4)$ contributions to
the potential.]

\subsec{Resummation: 2-loop diagrams}

\PSfig\figs{
  Resummation of \figp a.
}{9.5cm}{fig18.psx}{90}

Now let us turn to the resummation of actual diagrams.
The resummation of \figp a is depicted in \figs.

In the last pair of terms of \figs b, resumming the vector line makes no
difference because the $A_0$ polarization couples to the $k_0$
components of the scalar momenta, which are zero.  The only difference
between the last two diagrams is therefore the scalar masses.
It is easy to see by examining the general result \AVVa\ for the
setting-sun diagram, and remembering that the three-dimensional version
\Ithree\ of $I(m)$ is simply $I_3(m) = -mT/4\pi$, that these
three-dimensional diagrams do not depend on the scalar mass at
$O(e^4 T^4)$.  So their difference is ignorable and may be dropped.

In the other two pairs of terms of \figs b, the heavy loops simply
act as $O(e^2 T^2)$ mass insertions in the light loops.  For example,
the first pair of terms contributes to the potential
\eqn\bloom{
  \eqalign{
  &
    - {1\over2} e^2 T \intpthree
    \left[
       {n_\mu n_\nu \over p^2 + \ML^2} - {n_\mu n_\nu \over p^2 + M^2}
    \right]
    T \sum_{q_0\not=0} \intqthree
    {(P+2Q)_\mu (P+2Q)_\nu \over (Q^2+m^2)[(P+Q)^2+m^2]}
  \cr &
    = - {1\over2} e^2 T \intpthree
    \left[
       {n_\mu n_\nu \over p^2 + \ML^2} - {n_\mu n_\nu \over p^2 + M^2}
    \right]
    T \sum_{q_0\not=0} \intqthree
    {4 Q_\mu Q_\nu \over Q^4}
  \cr & \qquad\qquad\qquad\qquad \times
    \left[ 1 + O(p^2/Q^2, m^2/Q^2) + \cdots \right]
  \cr &
    = - {1\over4} T \intpthree
    \left[
       {1 \over p^2 + \ML^2} - {1 \over p^2 + M^2}
    \right]
    \times {-1\over3} e^2 T^2
    \left[ 1 + O(p^2/T^2, m^2/T^2) + \cdots \right]
  \cr &
    = - {1\over12\cdot4\pi} e^2 (\ML-M) T^3 + O(e^5 T^4),
  }
}
where $n_\mu = (1,\vec 0)$ and $P=(0,\vec p)$.
It is the $MT^3$ and $mT^3$ terms which resummation is supposed to
eliminate.  This does not occur on a graph-by-graph basis,
but they cancel in the sum over graphs.  The $\ML T^3$ and $\tilde mT^3$
terms cancel against the counter-term graphs of \figp h.
Evaluating the rest of \figs b similarly, one finds
\eqn\AVVaR{
     \mu^{2\eps} V^{\rm(a)} =
     \mu^{2\eps} V^{\rm(a)}_\naive
     - {1\over12\cdot4\pi} e^2 (\ML-M) T^3
     + O(e^5 T^4)
}

The results for the other diagrams may all be found in Appendix
\apndxResults.  When added together, they give the full, improved
2-loop potential:
\eqn\VAtwo{
  \eqalign{
     V^{(2)} =&
     {1\over2} \phi^2(T) \biggl\{
        - \mmu^2(T)
        + \left( {2\over3}\lambda(T) + 3e^2(T) \right) {T^2\over12}
  \cr & \qquad
        - {e^4 T^2\over(4\pi)^2} \left[
          2 \ln\left(3T\over2\ML\right)
          + 4\ln\left(3T\over2M\right)
          + {29\over9} - {2\over3} \cb - {3\over2}\cH
          \right] \biggl\}
  \cr &
     - {1\over12\pi} (2 M^3 + \ML^3) T
     + {1\over4!} \phi^4(T) \left[
        \lambda(T)
        + {36e^4\over(4\pi)^2}\left(\cb+{1\over3}\right) \right]
  \cr &
     + O(e^{9/2} T^4) ,
  }
}
where
\eqn\Arune{
  e^2(T) = e^2 + {e^4\over3(4\pi)^2} \ln{T^2\over\mub^2}
                        + O(e^5) ,
}
\eqn\Arunlam{
  \lambda(T) = \lambda + {18 e^4\over(4\pi)^2} \ln{T^2\over\mub^2}
                       + O(e^5) ,
}
\eqn\Arunphi{
  \phi(T) = \phi \left[ 1 + {3e^2\over2(4\pi)^2}\ln{T^2\over\mub^2}
                        + O(e^3) \right] ,
}
\eqn\Arunmmu{
  \mmu^2(T) = \mmu^2 + O(e^5 T^2) .
}
It is important to remember that this result was derived in the
high-temperature limit, and the $T \rightarrow 0$ limit of
the terms shown in \VAtwo\ is {\it not} the same as the zero-temperature
potential.  The $T \rightarrow 0$ limit of \VAtwo\ is trivial since
the $\lambda(T) \phi^4$ term vanishes if the scale of $\lambda$ is
run to zero.  For the true potential at zero temperature,
the relevant coupling is instead $\lambda(\phi)$.  This distinction is
important to keep in mind when using the 2-loop result for numerical
work.

\subsec{The vacuum shift}
\subseclab\secShift

How much do the 2-loop corrections just computed affect
the asymmetric vacuum at the phase transition?
The simplest quantity to examine is the vacuum expectation value.
This VEV is not a physical,
gauge-invariant quantity,\refmark\nielsen\
but the size of its shift due to 2-loop corrections will still
give us a good test of whether the loop corrections are small.
In a companion work,\refmark\magmass\
one of us has instead computed
the magnetic screening mass in the asymmetric vacuum, which is
physical and gauge invariant.  (Actually, we suspect that
the VEV may be gauge-invariant to the order we have computed,
but we do not know for sure.)

Normalize the potential so that $V(\phi,T)$ is zero at the origin.
The VEV $\phi$ of the asymmetric vacuum at the critical temperature $T$
is then determined by simultaneously solving
\eqn\VEVsolve{
  V(\phi,T) = 0, \qquad \partial_\phi V(\phi,T) = 0 .
}
Write $V = V_1 + \Delta V$ where $V$ is the full potential,
$V_1$ is some approximation (say, the one-loop potential), and
$\Delta V$ is a small perturbation.  Write $\phi = \phi_1 + \Delta\phi$
and $T = T_1 + \Delta T$ where $\phi_1$ is the asymmetric vacuum
of the approximation $V_1$ at its critical temperature $T_1$; that is,
\eqn\VEVsolveA{
  V_1(\phi_1,T_1) = 0, \qquad \partial_\phi V_1(\phi_1,T_1) = 0 .
}
Linearizing the equations \VEVsolve\ in $\Delta\phi$ and
$\Delta T$, one finds
\eqn\VEVsolnT{
  \Delta T \approx {-\Delta V \over \partial_T V_1} ,
}
\eqn\VEVsolnPhi{
  \Delta\phi \approx
     { \Delta V \partial_\phi\partial_T V_1
          - (\partial_\phi \Delta V) (\partial_T V_1)
       \over
       (\partial_\phi^2 V_1) (\partial_T V_1) } ,
}
where all the quantities are evaluated at the VEV.
Note that $\Delta\phi/\phi$ is order $\Delta V/V$, where $V$ is the
typical size of terms in the potential, and so $\Delta\phi/\phi$
is of order the loop expansion parameter.

Now identify $V_1$ with the one-loop potential and note that
\eqn\VoneDT{
  \partial_T V_1 = {1\over4} e^2 T \phi^2
           + O(e^3 T^2) .
}
Plugging into the solution \VEVsolnPhi\ for $\Delta\phi$:
\eqn\VEVsolnPhi{
  {\Delta\phi \over \phi} =
     - {\phi\over\tilde m_1^2} \partial_\phi
     \left(\Delta V\over \phi^2\right)
     + O(e^{3/2}) .
}
It is important when applying this formula to remember that
$\Delta V(\phi,T)$ should be replaced by
$\Delta V(\phi,T) - \Delta V(0,T)$ if the potential is not
already normalized to zero at $\phi=0$.
This result has the important property that is {\it vanishes} at
leading order if $\Delta V$ is proportional to $\phi^2$.
So the only
{\it two}-loop contributions to the potential \VAtwo\ which contribute
to the (leading-order) shift in the VEV are
those that involve logs of masses: the $e^4 T^2 \phi^2 \ln M(\phi)$ and
$e^4 T^2 \phi^2 \ln\ML(\phi)$ terms.
(The $e^4\phi^4$ terms, which would also contribute
if taken as part of $\Delta V$, came from the {\it one}-loop
contributions.)  Equation \VEVsolnPhi\ applies to all models
we shall examine.

For the Abelian Higgs model, \VEVsolnPhi\ reduces to
\eqn\VEVabelian{
  {\Delta\phi \over \phi} \approx
     {e^4 T^2\over(4\pi)^2\tilde m_1^2} \left(2 + {M^2 \over \ML^2}\right) ,
}
where $\ML$, $M$ and $\tilde m_1$
are as usual the effective masses at the
phase transition given by \Aml, \Amt\ and \goodm, and where the
one-loop approximation to the VEV $\phi$ is sufficient for the
right-hand side above.  The one-loop approximation to $\phi$ doesn't have
a simple form in terms of $T$, $\Mw(T=0)$, and $\mh(T=0)$
[at least not for the whole range $e^4 \ll \lambda \ll e^2$],
and is best computed numerically.

\newsec{Fermions}

Fermions are simpler to deal with than bosons because they do not
require resummation.  This is because Euclidean fermions have frequencies
$k_0 = (2n+1)\pi T$ which are never zero, and so their self-energies
may always be treated perturbatively.  For the same reason, the two-loop
contributions to the potential that involve fermions
will never have a logarithmic dependence on the fermion mass $\mf(\phi)$
and so will not affect the shift in the VEV.  Two-loop fermionic diagrams
do contribute to the effective potential at $O(g^4 T^4)$, however,
and so we shall treat them here (and later in our discussion of the
Minimal Standard Model) for completeness.

As a simple example of a theory analogous to the weak interactions,
let us chirally couple a single Dirac fermion to the Abelian Higgs
model:
\eqn\Lfermion{
  \eqalign{
     {\cal L} \rightarrow
  &
     {\cal L}
     + \bar\psi \left[ \dslash
          - i e \Lslash A \left(1-\gamma_5\over2\right) \right] \psi
     + \gy \bar\psi \left[ \Phi^* \left(1-\gamma_5\over2\right)
          + \Phi \left(1+\gamma_5\over2\right) \right] \psi
  \cr = &
     {\cal L}
     + \bar\psi \left[ \dslash
          - i e \Lslash A \left(1-\gamma_5\over2\right) \right] \psi
     + {\gy\over\sqrt2} \bar\psi (\phi_1 + i\gamma_5\phi_2) \psi .
  }
}
We shall treat the Yukawa coupling $\gy$ as being of the same order
as the gauge coupling $e$.
The fermion mass in the presence of a background field $\phi$ is
\eqn\Fmass{
   \mf = {1\over\sqrt2} \gy \phi ,
}
and the one-loop potential now picks up the familiar fermionic
contribution:
\eqn\VFone{
   \mu^{2\eps}V^{(1)} \rightarrow
   \mu^{2\eps}V^{(1)} + 4 \Jf(\mf) ,
}
where\refmark\dolan
\eqn\eqJf{
   \eqalign{
      \Jf(\mf) &=
      - {1\over2}\mu^{2\eps}\sumintf_K \ln(K^2+\mf^2)
   \cr &=
                 \const + {1\over48} \mf^2 T^2
                     + {1\over64\pi^2} \mf^4 \left[\Jepslogf\right]
   \cr & \qquad\qquad
                     + O(m^6/T^2) + O(\eps) ,
   }
}
and where we have defined the constant
\eqn\cfdef{
   \cf = \ln\pi - \gammaE = \cb - 2\ln2 .
}
The superscript (f) on the integral indicates that the frequency
sums are over fermionic momenta $k_0 = (2n+1)\pi T$.
The factor of four in \VFone\ corresponds to the four components of a
Dirac fermion.  As is conventional in \MSbar\ regularization, we
take the Dirac trace of one to be $\tr\,{\bf 1} = 4$ rather than
$2^{2-\eps}$.

Our working definition of $\gamma_5$ in dimensional regularization
is the naive one that\refmark\chanowitz
\eqn\gfivedefa{
   \{\gamma_5,\gamma_\mu\} = 0,
}
\eqn\gfivedefb{
   \gamma_5^2 = 1 .
}
This definition is adequate for calculations that do not require
traces involving an odd number of $\gamma_5$'s.
Additional properties are
\eqnn\gfivedefc
\eqnn\gfivedefd
$$
  \eqalignno{
      &\tr(\gamma_5\gamma_\mu\gamma_\nu\gamma_\sigma\gamma_\tau)
      \hbox{~is anti-symmetric;}
  & \gfivedefc \cr &
      \tr(\gamma_5\gamma_\mu\gamma_\nu\gamma_\sigma\gamma_\tau)
      = 4i\eps_{\mu\nu\sigma\tau} + O(\eps)\times{\rm ambiguity}
  \cr & \qquad\qquad\qquad
      \hbox{on the four-dimensional subspace
      $\mu,\nu,\sigma,\tau$ = 0,1,2,3.}
  & \gfivedefd \cr
  }
$$
The anomaly is intimately related to the fact that the ambiguous term
above cannot be explicitly defined.  In our calculation, however, this
term will never be relevant.

\PSfig\figt{
  Two-loop diagrams involving fermions.
}{4.3cm}{fig19.psx}{90}

The new 2-loop diagrams are shown in \figt.
As an example, take
\figt i and expand numerators in terms of denominators as we
did in \AVa\ through \AVVa\ for the Abelian Higgs model:
\eqn\FVx{
  \eqalign{
     \mu^{2\eps} V^{\rm(i)}_\naive &=
     - {1\over2} e^2 \sumintf_P \sumintf_Q
       { [\delta_{\mu\nu} - (P+Q)_\mu(P+Q)_\nu/(P+Q)^2]
         \over
         (P^2+\mf^2) (Q^2+\mf^2) [(P+Q)^2+M^2] }
  \cr & \qquad\qquad \times
       \tr (-i\Lslash P + \mf) \gamma_\mu \left(1-\gamma_5\over2\right)
         (i \Lslash Q + \mf) \gamma_\nu \left(1-\gamma_5\over2\right)
  \cr &=
     - {1\over8} e^2 \sumintf_P \sumintf_Q
       { [\delta_{\mu\nu} - (P+Q)_\mu(P+Q)_\nu/(P+Q)^2]
         \over
         (P^2+\mf^2) (Q^2+\mf^2) [(P+Q)^2+M^2] }
       \tr \Lslash P \gamma_\mu \Lslash Q \gamma_\nu
  \cr &=
     e^2 \bigl\{
       (1-\eps) [\If(\mf)]^2
       - 2(1-\eps) \If(\mf) I(M)
  \cr & \qquad\qquad
       + [(1-2\eps)\mf^2 - (1-\eps)M^2] \bar \Hff(\mf,\mf;M)
     \bigr\} .
  }
}
$\If$ and
$\bar\Hff$ are defined analogously to the bosonic $I$ and $\bar H$ of
previous sections:
\eqn\Ifdef{
   \If(\mf) = \mu^{2\eps} \sumintf_K {1 \over K^2+\mf^2} ,
}
\eqn\Hfdef{
   \bar \Hff(\mfa,\mfb; m)
   = \mu^{4\eps} \sumint_K \sumint_Q
           { 1
             \over
             (K^2+\mfa^2) (Q^2+\mfb^2) [(P+Q)^2+m^2] } .
}

The high-temperature expansion of $\If(\mf)$ may be obtained from
\eqJf\ using $\If = - \mf^{-1}\,\d J/\d \mf$:
\eqn\eqIf{
   \eqalign{
     \If(m) =
  &
           - {1\over24} T^2
           - {1\over16\pi^2} m^2 \left[\Jepslogf\right]
           + O(m^4/T^2)
   \cr&
           - \eps \cepsf {1\over 24} T^2 + O(\eps m^2) .
   }
}
In our earlier calculations, it turned out that the coefficient
$\ceps$ of the $\eps T^2$ term of $I(m)$ was unimportant because
it canceled in the final answer.  When fermions are added to the theory,
one finds that the final result {\it does} depend on the combination
$\cepsf-\ceps$, which it behooves us to calculate.
One may find by a derivation of $\If$ similar to that for $I$ in
\deriveI\ that
\eqn\eqcepsf{
   \cepsf = \ceps - 2\ln 2 .
}
We shall always make this substitution for $\cepsf$, and in this way
the $\ceps$'s will cancel in the final answer for the effective
potential.

The leading $O(T^2)$ contribution to $\bar \Hff$ turns out to
{\it vanish}, and
\eqn\eqHfbar{
   \bar \Hff(\mfa,\mfb; m) = O(mT) + O(\eps T^2).
}
A derivation is given in Appendix \apndxFermion.
The consequence of \eqHfbar\ is
that the $\bar \Hff$ term in \FVx\ may be ignored when studying the
effective potential through $O(e^4 T^4)$.
$\bar \Hff$ terms are similarly ignorable in all diagrams we
shall calculate.

The high-temperature expansion of our result \FVx\ for \figt i is
then
\eqn\FVVx{
  \eqalign{
     \mu^{2\eps} V^{\rm(i)}_\naive =
  &
     -{e^2 M T^3\over 12\cdot 4\pi} + {e^2 \mf^2 T^2\over 4(4\pi)^2} \left[
          {1\over\eps} +\ceps + \ln\left(\mub^2\over T^2\right)
          -1 - 2\cf - {2\over3} \ln 2        \right]
  \cr &
     - {e^2 M^2 T^2\over 12(4\pi)^2} \left[
          {1\over\eps} + \ceps + \ln\left(\mub^2\over T^2\right)
          - 1 - 2\cb - 2\ln 2 \right]
     + O(e^5 T^4) .
  }
}
The results for the rest of the diagrams, and the results of resumming
the $k_0=0$ modes of bosonic propagators, are given in Appendix
\apndxResults.  The final 2-loop potential is then
\eqn\FVtwo{
  \eqalign{
     V^{(2)} =&
     ~[\hbox{Abelian Higgs result of \VAtwo}]
  \cr &
     + {1\over2}\phi(T)^2 \biggl\{
         {1\over12}\gy^2(T) T^2
         + {\gy^4T^2\over(4\pi)^2} \left(
             - {1\over2} \cf - {1\over3}\ln2 \right)
  \cr & \qquad\qquad
         + {e^2 \gy^2 T^2\over(4\pi)^2} \left(
             - {1\over12} - {1\over2}\cf - {1\over6} \ln 2 \right)
         + {e^4 T^2\over (4\pi)^2} \left(
             {1\over18} + {1\over3}\cb + {1\over3}\ln 2 \right)
      \biggr\}
  \cr &
      + {1\over4!} \phi^4 \left[-{12\cf\gy^4\over(4\pi)^2}\right] ,
  }
}
where the running couplings $\lambda(T)$ and so forth in
\VAtwo\ now include fermionic effects.  (See Appendix \apndxResults\
for formulas for the running couplings.)

\newsec{Nonabelian Theories}

\PSfig\figu{
  Additional graphs needed for a non-abelian theory.
}{6.5cm}{fig20.psx}{80}

Extending our analysis to nonablelian theories is fairly straightforward;
it merely involves the computation of some new graphs, shown in \figu.
In Appendix \apndxResults, we give results for the SU(2) Higgs model
with a single Higgs doublet, defined by
\eqn\sutwoL{
  {\cal L} = -{1\over4} F^a_{\mu\nu} F^{a\mu\nu} + |D\Phi|^2 - V(|\Phi|^2) ,
  \qquad
  V(|\Phi|^2) = -\mmu^2 |\Phi|^2 + {1\over3!}\lambda |\Phi|^4 ,
}
where $\Phi$ is an SU(2) doublet and
\eqn\coderiv{
  D_\mu\Phi = \left(\partial_\mu - i {1\over2} g A_\mu \cdot \tau \right) \Phi.
}
We once again normalize $\phi$ to $\phi^2 = 2|\Phi|^2$.

Reducing the graphs of \figu\ to simple scalar integrals like $I(m)$
and $\bar H(m_1,m_2,m_3)$ turns out to require the introduction of
a new function:
\eqn\defL{
   L(m_1,m_2) = \sumint_P \sumint_Q {
                   (P \cdot Q)^2
                   \over
                   P^2 (P^2+m_1^2) Q^2 (Q^2+m_2^2)
                 } .
}
One can derive a high-temperature expansion of this function, which we
give in Appendix \apndxFermion, but it
is unnecessary because terms involving $L(m_1,m_2)$ cancel exactly
between figs.~\xfig\figu m and \xfig\figu o.

\PSfig\figv{
  Resummation of \figu m.
}{8cm}{fig21.psx}{90}

\PSfig\figw{
  Rewriting of \figv.  The L and T lines represent the longitudinal ($A_0$)
  and transverse polarizations of {\it only} the $p_0=0$ contributions.
}{8cm}{fig22.psx}{90}

The only part of the calculation of the two-loop potential that is not
a completely straightforward extension of our previous model calculations
is the resummation of \figu m.
The resummation is shown in \figv.
Recall that only the longitudinal
($A_0$) polarization is resummed, and split propagators into
longitudinal (L) and transverse (T) parts.
By using the fact
that the three-vector coupling vanishes for $LLL$ and $LTT$ when all
frequencies $p_0$ are zero,
\figv\ may be rewritten
as \figw.
The three dimensional diagram in the last pair of terms is
easy to evaluate, and the result is given in Appendix \apndxResults.

The total ring-improved 2-loop potential is given by
\eqn\SUVtwo{
  \eqalign{
    V^{(2)} =
  &
    {1\over2} \phi^2(T) \biggl\{
      - \mmu^2(T)
      + \left[ \lambda(T) + {9\over4} g^2(T) \right] {T^2\over12}
  \cr & \qquad
      + {g^4 T^2\over(4\pi)^2} \biggl[
        {107\over24} + {5\over4}\cb - {81\over64}\cH
        + {9\over4}\ln\left(3T\over 2\ML + M\right)
  \cr & \qquad\qquad
        + {63\over16}\ln{T\over M}
        - {3\over8}\ln\left(3T\over2\ML\right)
        - {3\over4}\ln\left(3T\over2M\right)
      \biggr] \biggr\}
  \cr &
    + {3g^2T^2\over(4\pi)^2} \left[ \ML M - 2 \Mdebye^2
                                  \ln\left(2\ML+M\over2\Mdebye\right) \right]
  \cr &
    - {T\over12\pi} (6M^3 + 3\ML^3)
    + {1\over4!}\phi^4(T) \left[ \lambda(T)
         + {27g^4\over4(4\pi)^2}\left(\cb+{1\over3}\right) \right] ,
  }
}
where we have introduced the notation
\eqn\defMdebye{
  \Mdebye^2 = \ML^2 - M^2}
for the $\phi$-independent thermal mass.  Note that though the $\ML M$
term above is linear in $\phi$ as $\phi \rightarrow 0$, this linear
term cancels against the $\phi\rightarrow 0$ behavior of the
$\Mdebye^2 \ln(2\ML+M)$ term.

We can now compute the relative shift in the VEV from its one-loop
value using \VEVsolnPhi:
\eqn\VEVsu{
    {\Delta\phi\over\phi} \approx
    {g^4 T^2\over(4\pi)^2 \tilde m_1^2} \biggl[
      {87\over32} + {3 \Mdebye^2\over 2\ML M}
      - {3M^2\over16\ML^2}
      - {3\Mdebye^2\over M^2} \ln\left(2\ML+M\over 2\Mdebye\right)
      \biggr] ,
}
where $M$, $\ML$, and $\tilde m_1$ may be evaluated at the phase transition
using the one-loop approximation.
$\tilde m_1$ is obtained by taking the second derivative of the
one-loop potential through $O(g^3 T^2)$:
\eqn\SUmtilde{
  \tilde m_1^2 = m_1^2 + \left(\lambda+{9\over4}g^2\right){T^2\over12}
               - {3 g^2 M T\over4\pi}
               - {3 g^2 T\over16\pi} \left({M^2\over\ML}+\ML\right) .
}
Note that the $\ML M$ term in the potential contributes to $\Delta\phi/\phi$,
whereas only logarithms contributed in the Abelian case.

\newsec{The Minimal Standard Model}

Finally, consider the Minimal Standard Model with a single Higgs doublet,
which differs from the previous models only by the complexity of keeping
track of the various couplings.  Our conventions for coupling constants
are such that
\eqn\coderivb{
  D_\mu = \partial_\mu + {1\over2} g_2 A_\mu \cdot \tau
           + {1\over2} Y g_1 B_\mu ,
}
the hypercharge is normalized so that $Q=T_3+Y/2$, and the Yukawa
coupling is
\eqn\yukawa{
  \gy \,\bar q_{\rm L} \cdot \Phi \, t_{\rm R} + \hbox{h.c.}
}
where $\Phi$ is the full complex doublet.  $\gy$ is the top quark
coupling, which is the only Yukawa coupling that we treat as
non-zero.  We shall also include QCD interactions for the quarks,
with coupling $\gs$.  The couplings $g_1$, $g_2$, $\gy$, and $\gs$
are formally treated as all having the same order of magnitude $g$.
$\nf$ represents the number of families and is three in the
Minimal Standard Model.

At zero temperature, the mass matrix for the Z boson and the photon,
in the $(A^{(3)}, B)$ basis, is
\eqn\mixzero{
   M^2 = \pmatrix{
           {1\over4}g_2^2\phi^2  & -{1\over4}g_1 g_2 \phi^2 \cr
          -{1\over4}g_1g_2\phi^2 &  {1\over4}g_1^2 \phi^2
         }
   =
   \pmatrix{
      \cos\thetaw  &  \sin\thetaw \cr
     -\sin\thetaw  &  \cos\thetaw
   }
   \pmatrix{
      \Mz^2 & ~ \cr ~ & 0
   }
   \pmatrix{
      \cos\thetaw  & -\sin\thetaw \cr
      \sin\thetaw  &  \cos\thetaw
   } .
}
When we include the Debye masses generated for the longitudinal ($A_0$)
components at finite temperature however, one finds\refmark\carrington
\eqn\mixT{
  \eqalign{
    \ML^2 =
  &
    \pmatrix{
           {1\over4}g_2^2\phi^2 + \left({5\over6}+{\nf\over3}\right)g_2^2T^2
       &  -{1\over4}g_1 g_2 \phi^2
       \cr-{1\over4}g_1g_2\phi^2
       &   {1\over4}g_1^2\phi^2 + \left({1\over6}+{5\nf\over9}\right)g_1^2T^2
    }
    \equiv
  \cr & \qquad\qquad
    \pmatrix{
       \cos\thetat  &  \sin\thetat \cr
      -\sin\thetat  &  \cos\thetat
    }
    \pmatrix{
       \MLz^2 & ~ \cr ~ & \MLg^2
    }
    \pmatrix{
       \cos\thetat  & -\sin\thetat \cr
       \sin\thetat  &  \cos\thetat
    } ,
  }
}
and we have diagonalized to define an effective mixing angle
$\thetat(\phi,T)$ and effective longitudinal masses
$\MLz(\phi,T)$ and $\MLg(\phi,T)$.

\PSfig\figx{
  2-loop graphs for the Minimal Standard Model.  ``q'' indicates
  all quarks other than the top quark.
}{20.5cm}{fig23.psx}{90}

The result for the two-loop potential is given in Appendix
\apndxResults, where we list the contribution from each of the
graphs of \figx.
Only those diagrams which give a
$\phi$-dependent contribution to the potential at $O(g^4 T^4)$
are shown.  Note that there is a QCD correction in
\figx i$_1$ which gives a potentially large contribution to the
potential of order $\gs^2 \gy^2 \phi^2 T^2$ but which, like all
fermionic contributions, does not contribute to the leading-order
shift in the VEV from its one-loop value.

It is cumbersome to write down the total answer for the two-loop
potential, and we shall leave it in the form of Appendix
\apndxResults\ where each diagram is given separately.
The reader should note that the relative shift in the VEV
is given by \VEVsu\ in the limit that $g_1\rightarrow0$.  This is because,
as discussed earlier, the fermion contributions do not affect the
shift in the VEV except indirectly through their effect on the
one-loop values of $M$, $\ML$, and $\tilde m_1$ at the phase transition.
For example, $\tilde m_1$ in this limit changes from \SUmtilde\ to
\eqn\MSMtildemLim{
  \tilde m_1^2 = m_1^2
               + \left(\lambda+{9\over4}g^2+3\gy^2\right){T^2\over12}
               - {3 g^2 M T\over4\pi}
               - {3 g^2 T\over16\pi} \left({M^2\over\ML}+\ML\right) .
}

\newsec{Numerical Results}

As a test of the size of loop corrections, we shall now compute the
shift $\Delta\phi$ in the VEV at the phase transition due to the inclusion of
two-loop corrections in the Minimal Standard Model.
We do this numerically, by running couplings
to $\mub = T$ and then finding the phase transition for both the one-loop
and two-loop approximations to the potential given in Appendix
\apndxResults.

As discussed in section \secShift, the relative
size $\Delta\phi/\phi$ of this shift is $O(g)$, the loop expansion
parameter.  This means that we need to specify the original
zero-temperature parameters of the theory, such as $\lambda$ and $g^2$,
to relative order $g$ in terms of physical quantities such as the
zero-temperature Higgs and vector masses.  Tree-level relations are
adequate for most quantities, since corrections at zero-temperature
are suppressed by $g^2$ instead of $g$.
The exceptions are $\lambda$ and $\mmu^2$
since they are
$O(g^3)$ by our power-counting convention and
receive $O(g^4)$ corrections at one-loop.
To obtain $\lambda$ and $\mmu^2$ in terms of physical masses, consider
the one-loop potential at zero-temperature:
\eqn\VoneZero{
  \eqalign{
     V^{(1)}(\phi, T=0) =
  &
     - {1\over2} \mmu^2\phi^2 + {1\over4!} \lambda\phi^4
  \cr &
     + {6\Mw^4(\phi)\over64\pi^2}
           \left[\ln\left(\Mw^2(\phi)\over\mub^2\right)-{5\over6}\right]
     + {3\Mz^4(\phi)\over64\pi^2}
           \left[\ln\left(\Mz^2(\phi)\over\mub^2\right)-{5\over6}\right]
  \cr &
     - {3\mt^4(\phi)\over16\pi^2}
           \left[\ln\left(\mt^2(\phi)\over\mub^2\right)-{3\over2}\right]
     + O(m_1^4, m_2^4) .
  }
}
We shall refer to the zero-temperature VEV as $\sigma$.  The
physical Higgs mass at the order under consideration is just
the second derivative of this potential at $\sigma$.\foot{
  The physical Higgs mass is actually determined by the pole of the
  Higgs propagator and so is given by the solution to the real-time
  dispersion relation $P^2 - M^2 = \Pi(P^2)$ where $\Pi$ is the
  self-energy.  The claim that it is given by the second derivative of
  the effective potential corresponds to the approximation
  $\Pi(P^2) \rightarrow \Pi(0)$ in this dispersion relation.
  The difference $\Pi(P^2)-\Pi(0)$ is order $g^5\sigma^2$ and
  does not affect our derivation of $\lambda$ through $O(g^4)$.
}
Solving for $\lambda$ and $\mmu^2$ in terms of the physical masses,
one finds
\eqn\LamZero{
  \eqalign{
     \lambda = {3\bar\mh^2\over\sigma^2}
       - {3\over32\pi^2} \bigl\{
          {3\over2} g_2^4
             \left[\ln\left(\bar\Mw^2\over\mub^2\right) + {2\over3}\right]
  &
          + {3\over4} (g_1^2+g_2^2)^2
             \left[\ln\left(\bar\Mz^2\over\mub^2\right) + {2\over3}\right]
  \cr & \qquad
          - 12 \gy^4 \ln\left(\bar\mt^2\over\mub^2\right)
         \biggr\}
       + O(g^5) ,
  }
}
\eqn\MuZero{
   \mmu^2 = {1\over2} \bar\mh^2
     - {\sigma^2\over64\pi^2} \left[
        {3\over2} g_2^4 + {3\over4}(g_1^2+g_2^2)^2 - 12\gy^4 \right]
     + O(g^2 \bar\mh^2) ,
}
where the bars denote that masses are evaluated at the VEV $\sigma$
at zero temperature with renormalization scale $\mub$,
and where $\sigma$ may be expressed as $\sigma = \bar\Mw/g$.

\PSfig\figy{
   Relative shift $\Delta\phi/\phi$ of the asymmetric vacuum at
   the critical temperature plotted vs. the Higgs mass.
   The solid line comes from using the full formula for the 2-loop
   potential derived in this paper; the dashed line should approximately
   match it up to yet high-order corrections in the loop expansion.
   The high-temperature expansion used in this paper is valid provided
   the dotted line (described in the text) is small compared to one.
}{10cm}{fig24.psx}{75}

Our numerical results for the shift $\Delta\phi/\phi$ due to 2-loop
corrections are shown in \figy\
as a function of $\mh$, assuming a top quark mass of $100$ GeV.
To get the solid line, we computed $\phi$ and $\Tc$ independently
for both the one-loop potential and for our full result for the
2-loop potential.  To help control errors of the one-loop computation
at very small $\mh$, we have used exact results for the one-loop
fermion and vector contributions rather than high-temperature
expansions.

Alternatively, the dashed line gives the simpler
calculation of the pure SU(2) result \VEVsu\ for the shift, which
we have evaluated using the Minimal Standard Model results
for $\Mw$ and $\MLw$ in place of $M$ and $\ML$.  We have also
replaced $\tilde m_1^2$ by the complete second-derivative of the
one-loop potential rather than by \MSMtildemLim\ because,
due to fine cancellations among terms, the assumption that
$g_1\rightarrow 0$ used in \MSMtildemLim\ turns out to be a
very bad approximation if the one-loop VEV and critical temperature
were computed with $g_1 \not= 0$.
The dashed line should approximately match the solid line when both
(1) $\Delta\phi/\phi$ is small and (2) $\mh$ is not so small that the
high-temperature expansion has broken down.  This correspondence
is evident in \figy.
The deviation of the lines at large $\mh$ is a manifestation of
the breakdown of the loop expansion.

The dotted line in \figy\ is a diagnostic of our ubiquitous
assumption ($\lambda\gg g^4$) that we may use the high-temperature
expansion.  It is the ratio of the $O(M^6/T^2)$
piece of the one-loop potential to the $O(M^4)$ piece.
This ratio is taken in the asymmetric vacuum
at the phase transition.
More specifically,
\eqn\CheckExpand{
  \eqalign{
     v_4 &= {1\over32\pi^2} \left( 6\cb\Mw^4 + 3\cb\Mz^4 - 12\cf\mt^4 \right),
  \cr
     v_6 &= {\zeta(3)\over768\pi^4T^2}
               \left( 6\Mw^6 + 3\Mz^6 - 7\cdot12\mt^6 \right),
  }
}
and the dotted line is $|v_6/v_4|$.  One can see in \figy\ that the
high-temperature expansion begins to break down at small $\mh$.

\PSfig\figyb{
  The ratio $\phi/T$ at the phase transition.  The dashed line is the
  one-loop result and the solid line is our two-loop result in Landau
  gauge.
}{10.5cm}{fig25.psx}{75}

Fig.~\xfig\figyb\ shows the result for $\phi/T$ at the phase transition, which
is the quantity used in Ref.~\dinebound\ to extract the upper
bound on the Higgs mass for weak baryogenesis.  According to
Ref.~\dinebound, a necessary requirement for a successful
scenario of baryogenesis is that $\phi/T \ge 1$.  The dashed and
solid lines show our one-loop and two-loop results respectively.
At the experimental lower bound of 60 GeV, the one-loop result of
0.47 is inadequate.  The 2-loop corrections boost $\phi/T$ by 40\%
to roughly 0.65.  This remains inadequate.  Unfortunately,
the corrections are large enough that this conclusion does not
impress us as airtight---the validity of the loop expansion is
only marginal here.
Note that, for a consistent comparison at this order, the sphaleron
mass and resulting limit on $\phi/T$ should also be consistently
determined to the same order.

\PSfig\figyc{
  Same as \figy\ but for $\mt$=180 GeV.
}{10.5cm}{fig26.psx}{75}

\PSfig\figyd{
  Same as \figyb\ but for $\mt$=180 GeV.
}{10.5cm}{fig27.psx}{75}

Figs.~\xfig\figyc\ and \figyd\ show our results for a top mass of
180 GeV.
In parting, we remind the reader that the Minimal Standard Model is
only a specific testing ground for these issues and that the particular
constraints just discussed are evaded in multiple-Higgs models.

\vskip 0.5in

Both P.A. and O.E. were supported in part by the
U.S. Department of Energy, grant DE-FG06-91ER40614.
P.A. was also supported in part
by the Texas National Research Laboratory Commission.
O.E. was also supported in part by a Fundaci\'on Andes fellowship,
program C-12021/7.
We are indebted to Larry Yaffe for many long and useful discussions
and to Lowell Brown for the derivation of $\bar H(m_1,m_2,m_3)$ in
three dimensions.

\def\skipbit{\vskip 0.1in}

\appendix{A}{Results for 2-Loop diagrams}

In this appendix we collect 2-loop results for individual diagrams in the
various theories discussed in this paper.  Since the theories involve many
of the same diagrams, and differ only by group factors, we only compute
them once.  The ${\cal D}$ functions defined at the end of this
appendix are the results of diagrams where we have factored out
a combination of coupling constants, symmetry factors, and group factors.
The subscripts of the ${\cal D}$'s indicate the types of particles
involved in the diagram: S for scalar, V for vector, F for fermion, and
$\eta$ for ghost.  Note that that the effective potential and the sum
over diagrams have a relative minus sign in Euclidean space.
We shall display only the $\phi$-dependent pieces of contributions
to the potential and shall not explicitly indicate the presence of
temperature-dependent constants by ``$\const$''
For brevity, we shall drop the unimportant $\mu^{2\eps}$ terms that
keep track of the dimensions of the potential $V$ and the field $\phi$;
but every instance of $V$ and $\phi$ in this appendix should be
understood as $\mu^{2\eps} V$ and $\mu^\eps \phi$.
Throughout this appendix we adopt the notation
\eqn\epsbar{
   {1\over\bar\eps} = {1\over\eps} (1+\eps\ceps) ,
}
where $\ceps$ is the unimportant constant of \eqI, and we define
\eqn\Hthree{
   \bar H_3(m_1,m_2,m_3)
   = - {1\over(4\pi)^2} T^2 \ln(m_1+m_2+m_3),
}
to be the (non-constant) three-dimensional piece of $\bar H$, as in
\Hdiffb.

To actually use the formulas of this appendix
for computing two-loop corrections
numerically, sum the individual contributions but
(1) drop all $1/\bar\eps$ terms, since they cancel;
(2) set $\mub$ to $T$ and so drop all $\ln(\mub^2/T^2)$ terms;
and so (3) replace couplings $g$,
masses $\mmu$ and fields $\phi$ everywhere by $g(T)$, $\mmu(T)$
and $\phi(T)$.

Unresummed results for the diagrams in (A.1) and (A.2) may be obtained
by replacing all ${\cal D}$'s by ${\cal D}^\naive$'s and dropping
$V^{\rm (h)}$.

\subsec{Abelian Higgs Model}
\subseclab\Aseca

\noindent
Counter terms:
\eqn\AAcounter{
  \matrix{
    \phi_\bare
       &= \left(1 + {3e^2\over2(4\pi)^2\eps} +\cdots\right)\phi
  \qquad &
    A_\bare
       &= \left(1 - {e^2\over6(4\pi)^2\eps} + \cdots\right) A
  \cr
    e_\bare^2
       &= \left(1 + {e^2\over3(4\pi)^2\eps} + \cdots\right) e^2\mu^{2\eps}
  \qquad &
    \lambda_\bare
       &= \left(\lambda + {18e^4\over(4\pi)^2\eps} + \cdots\right) \mu^{2\eps}
  \cr
    \mmu_\bare^2
       &= \left(1 - {3e^2\over(4\pi)^2\eps} + \cdots\right) \mmu^2
  \qquad &
    ~
  \cr
  }
}
Multiplicative corrections are shown only through
order $e^2$, and order $\lambda$ corrections are left out since
$\lambda \sim e^3$ by assumption.  Running couplings and fields may
be found by replacing $1/\eps \rightarrow \ln(T^2/\mub^2)$ and
$\phi_\bare\rightarrow\phi(T)$, $e^2_\bare\rightarrow e^2(T)$,
\etc\ above.
\skipbit
\noindent
Masses:
\eqn\AAmassa{
      M^2 = e^2\phi^2 ,
   \qquad
      m_1^2 = -\mmu^2 + {1\over2}\lambda\phi^2 ,
   \qquad
      m_2^2 = -\mmu^2 + {1\over6}\lambda\phi^2
}
\eqn\AAmassb{
      \ML^2 = e^2\phi^2 + {1\over3} e^2 T^2 \hfil ,
   \qquad\qquad
      \tilde m_i^2 = m_i^2
                + \left( {2\over3}\lambda+3e^2\right)
                  {T^2\over12}
                + \cdots
}
We shall not bother showing the sub-leading corrections to the
thermal scalar mass, as in \goodm, since they cancel out in the
final result for the potential to $O(e^4 T^4)$.
\skipbit
\noindent
1-loop result: See eq.~\AVoneR.
\skipbit
\noindent
2-loop diagrams of \figp:
\eqna\AAdiags
$$
 \eqalignno{
   V^{\rm(a)}
        &	= - {1\over2} e^2 {\cal D}_{\rm SSV}(m_1,m_2,M) ,
 & \AAdiags a \cr
   V^{\rm(b)}
        &= - {1\over4} e^4 \phi^2 {\cal D}_{\rm SVV}(m_1,M,M)
 & \AAdiags b \cr
   V^{\rm(c)}
        &= - {1\over4} e^2 [
              {\cal D}_{\rm SV}(m_1,M) + {\cal D}_{\rm SV}(m_2,M) ]
 & \AAdiags c \cr
   V^{\rm(d)}
        &= {e^2 M^2 T^2 \over 24 (4\pi)^2}
              \left({1\over\bar\eps} - {2\over3}\right)
        + O(e^5 T^4)
 & \AAdiags d \cr
   V^{\rm(e)}
        &= {3 e^4 \phi^2 T^2 \over 8 (4\pi)^2}
              \left({1\over\bar\eps} - {2\over3}\right)
        + O(e^5T^4)
 & \AAdiags e \cr
   V^{\rm(f)}
        &= O(e^5T^4)
 & \AAdiags f \cr
   V^{\rm(g)}
        &= {e^4\phi^2 T^2 \over 2(4\pi)^2} {1\over\bar\eps}
        + O(e^{11/2}T^4)
 & \AAdiags g \cr
   V^{\rm(h)}
        &= {e^2 \ML T^3\over 24\pi}
        + {e^2 (\tilde m_1 + \tilde m_2) T^3\over 32\pi}
        + O(e^{9/2}T^4)
 & \AAdiags h \cr
 }
$$

\subsec{Chiral Abelian Higgs Model}

\noindent
Counter terms:
\eqn\AFcounter{
  \matrix{
    \phi_\bare
       &= \left(1 + {(3e^2-\gy^2)\over2(4\pi)^2\eps}
                +\cdots \right)\phi
  \qquad &
    A_\bare
       &= \left(1 - {e^2\over2(4\pi)^2\eps}
                + \cdots\right) A
  \cr
    \psi_{{\rm L},\bare}
       &= \left(1 - {\gy^2\over4(4\pi)^2\eps}
                + \cdots\right) \psi_{\rm L}
  \qquad &
    \psi_{{\rm R},\bare}
       &= \left(1 - {\gy^2\over4(4\pi)^2\eps}
                + \cdots\right) \psi_{\rm R}
  \cr
    e_\bare^2
       &= \left(1 + {e^2\over(4\pi)^2\eps} + \cdots\right) e^2 \mu^{2\eps}
  \qquad &
    \lambda_\bare
       &= \left(\lambda + {(18e^4-6\gy^4)\over(4\pi)^2\eps} + \cdots\right)
          \mu^{2\eps}
  \cr
    \mmu_\bare^2
       &= \left(1 - {(3e^2-\gy^2)\over(4\pi)^2\eps} + \cdots\right) \mmu^2
  \qquad &
    g_{{\rm y},\bare}
       &= \left(1 + {(2\gy^2-3e^2)\over2(4\pi)^2\eps} + \cdots\right) \gy
          \mu^{\eps}
  \cr
  }
}
Formulas for running couplings may be obtained as in section \Aseca.
\skipbit
\noindent
Masses are as \AAmassa\ with:
\eqn\AFmassb{
      \mf^2 = {1\over2} \gy^2 \phi^2 ,
   \qquad
      \ML^2 = e^2\phi^2 + {1\over2} e^2 T^2 \hfil ,
   \qquad
      \tilde m_i^2 = m_i^2
                + \left( {2\over3}\lambda+3e^2+\gy^2\right)
                  {T^2\over12}
                + \cdots .
}
\skipbit
\noindent
1-loop result:
\eqn\AFone{
  \eqalign{
     V^{(1)}
     = {1\over2} & \left[ -\mmu^2
         + \left({2\over3}\lambda+3e^2+\gy^2\right) {T^2\over12} \right] \phi^2
     - {1\over12\pi} (2M^3 + \ML^3) T
     + {1\over4!}\lambda\phi^4
  \cr&
     - {3 M^4\over 4(4\pi)^2} \left[
         \ln\left(\mub^2\over T^2\right) - {2\over3} - 2\cb \right]
     + {\mf^4\over (4\pi)^2} \left[
         \ln\left(\mub^2\over T^2\right) - 2\cf \right]
  \cr&
     + O(e^{9/2} T^4)
  }
}
\skipbit
\noindent
2-loop diagrams of figs.~\xfig\figp a--c same as \AAdiags.

\noindent
2-loop diagrams of figs.~\xfig\figp d--h:
\eqna\AFdiags
$$
 \eqalignno{
   V^{\rm(d)}
        &= {e^2 M^2 T^2 \over 8 (4\pi)^2}
              \left({1\over\bar\eps} - {2\over3}\right)
        + O(e^5 T^4)
 & \AFdiags d \cr
   V^{\rm(e)}
        &= {e^2 (3e^2-\gy^2) \phi^2 T^2 \over 8 (4\pi)^2}
              \left({1\over\bar\eps} - {2\over3}\right)
        + O(e^5T^4)
 & \AFdiags e \cr
   V^{\rm(f)}
        &= O(e^5T^4)
 & \AFdiags f \cr
   V^{\rm(g)}
        &= {(3e^4-\gy^4)\phi^2 T^2 \over 6(4\pi)^2} {1\over\bar\eps}
        + O(e^{11/2}T^4)
 & \AFdiags g \cr
   V^{\rm(h)}
        &= {e^2 \ML T^3\over 16\pi}
        + {(3e^2+\gy^2) \over 96\pi} (\tilde m_1 + \tilde m_2) T^3
        + O(e^{9/2}T^4)
 & \AFdiags h \cr
 }
$$

2-loop diagrams of \figt:
$$
 \eqalignno{
   V^{\rm(i)}
        &= - {1\over2} e^2 {\cal D}_{\rm FFV}(\mf,\mf,M) ,
 & \AFdiags i \cr
   V^{\rm(j)}
        &= - {1\over4} \gy^2 {\cal D}_{\rm FFS}(\mf,\mf,m_1)
           - {1\over4} \gy^2 {\cal D}_{\rm FFS}(\mf,\mf,m_2)
 & \AFdiags j \cr
   V^{\rm(k)}
        &= {\gy^2\mf^2 T^2\over12(4\pi)^2}
           \left({1\over\bar\eps} - 2\ln2 \right)
           + O(e^6 T^4)
 & \AFdiags k \cr
   V^{\rm(l)}
        &= 0
 & \AFdiags l \cr
 }
$$

\subsec{Gauged SU(2) Higgs Theory}

\noindent
Take
$\gs=\gy=g_1=e=\thetaw=\thetat=0$; set $\Mz=\Mw=M$, $\MLz=\MLw=\ML$,
$g_2=g$; and ignore $\MLg$
in the Minimal Standard Model results below.

\subsec{Minimal Standard Model}

\noindent
Counter terms:
\eqn\AFcounter{
  \eqalign{
    \phi_\bare
       &= \left(1 + {(9g_2^2+3g_1^2-12\gy^2)\over8(4\pi)^2\eps}
                +\cdots \right)\phi
  \cr
    A_\bare
       &= \left[1
            + \left({25\over12}-{2\nf\over3}\right){g_2^2\over(4\pi)^2\eps}
            + \cdots\right] A
  \cr
    B_\bare
       &= \left[1
            - \left({1\over12}+{10\nf\over9}\right){g_1^2\over(4\pi)^2\eps}
            + \cdots\right] B
  \cr
    t_{{\rm L},\bare}
       &= \left(1 - {\gy^2\over4(4\pi)^2\eps} +\cdots\right) \psi_{\rm L}
  \cr
    t_{{\rm R},\bare}
       &= \left(1 - {\gy^2\over2(4\pi)^2\eps} +\cdots\right) \psi_{\rm R}
  \cr
    g_{2,\bare}^2
       &= \left[1
        - \left({43\over6}-{4\nf\over3}\right){g_2^2\over(4\pi)^2\eps}
        + \cdots\right] g_2^2 \mu^{2\eps}
  \cr
    g_{1,\bare}^2
       &= \left[1
        + \left({1\over6}+{20\nf\over9}\right){g_1^2\over(4\pi)^2\eps}
        + \cdots\right] g_1^2 \mu^{2\eps}
  \cr
    g_{{\rm y},\bare}
       &= \left[1
        + \left({9\over4}\gy^2-4\gs^2-{9\over8}g_2^2-{17\over24}g_1^2\right)
                    {1\over(4\pi)^2\eps}
        + \cdots\right] \gy \mu^\eps
  \cr
    \lambda_\bare
       &= \left[1 + \left( {27\over8}g_2^4 + {9\over4}g_2^2g_1^2
                           + {9\over8}g_1^4 - 18\gy^4 \right)
                    {1\over(4\pi)^2\eps}
                  + \cdots\right] \lambda \mu^{2\eps}
  \cr
    \mmu_\bare^2
       &= \left(1 - {(9g_2^2+3g_1^2-12\gy^2)\over4(4\pi)^2\eps}
                +\cdots \right)\mmu^2
  }
}
Formulas for running couplings may be obtained as in section \Aseca.
\skipbit
\noindent
Masses:
\eqn\AMSMmassa{
      \Mw^2 = {1\over4} g_2^2\phi^2 ,
   \qquad
      \Mz^2 = {1\over4} (g_2^2+g_1^2)\phi^2 ,
   \qquad
      \mf^2 = {1\over2} \gy^2 \phi^2
}
\eqn\AMSmassb{
      m_1^2 = -\mmu^2 + {1\over2}\lambda\phi^2 ,
   \qquad
      m_2^2 = -\mmu^2 + {1\over6}\lambda\phi^2
}
\eqn\AMSmassc{
      \MLw^2 = {1\over4} g_2^2\phi^2
             + \left({5\over6}+{\nf\over3}\right)g_2^2 T^2
}
\eqn\AMSmassd{
      \tilde m_i^2 = m_i^2
                + \left( \lambda+{9\over4}g_2^2+{3\over4}g_1^2+3\gy^2\right)
                  {T^2\over12}
                + \cdots .
}
\eqn\mixTb{
  \eqalign{
  &
    \pmatrix{
           {1\over4}g_2^2\phi^2 + \left({5\over6}+{\nf\over3}\right)g_2^2T^2
       &  -{1\over4}g_1 g_2 \phi^2
       \cr-{1\over4}g_1g_2\phi^2
       &   {1\over4}g_1^2\phi^2 + \left({1\over6}+{5\nf\over9}\right)g_1^2T^2
    }
    \equiv
  \cr & \qquad\qquad\qquad\qquad\qquad
    \pmatrix{
       \cos\thetat  &  \sin\thetat \cr
      -\sin\thetat  &  \cos\thetat
    }
    \pmatrix{
       \MLz^2 & ~ \cr ~ & \MLg^2
    }
    \pmatrix{
       \cos\thetat  & -\sin\thetat \cr
       \sin\thetat  &  \cos\thetat
    } ,
  }
}
\skipbit
\noindent
1-loop result:
\eqn\AMSMone{
  \eqalign{
     V^{(1)}
     = {1\over2} & \left[ -\mmu^2
         + \left(\lambda+{9\over4}g_2^2+{3\over4}g_1^2+3\gy^2\right)
         {T^2\over12} \right] \phi^2
  \cr&
     - {1\over12\pi} (4\Mw^3 + 2\Mz^3 + 2\MLw^3 + \MLz^3 + \MLg^3) T
     + {1\over4!} \lambda\phi^4
  \cr&
     - {3 (2\Mw^4+\Mz^4)\over 4(4\pi)^2} \left[
         \ln\left(\mub^2\over T^2\right) - {2\over3} - 2\cb \right]
     + {3\mf^4\over (4\pi)^2} \left[
         \ln\left(\mub^2\over T^2\right) - 2\cf \right]
  \cr&
     + O(g^{9/2} T^4)
  }
}
\skipbit
\noindent
2-loop diagrams of \figx:

\noindent
$\cos\thetaw$ is defined below by the tree-level relation
$\tan\thetaw = g_1/g_2$, and $e$ is $g_2 \sin\thetaw$.
We shall also use the short-hand notation
\eqn\Mshort{
  \eqalign{
    \dMa &= 2\MLw + \MLz \cos^2\thetat + \MLg \sin^2\thetat
            - 2\Mw - \Mz \cos^2\thetaw ,
  \cr
    \dMb &= \MLz \sin^2\thetat + \MLg \cos^2\thetat - \Mz \sin^2\thetaw .
  }
}
Take care to note that many of the graphs below are written in terms of the
${\cal D}^\naive$ instead of the ${\cal D}$.  This is because of notational
complications caused by the difference between $\thetaw$ and $\thetat$.
\eqna\AMSMdiags
$$
 \eqalignno{
   V^{\rm(a)}
      & = - {1\over4} g_2^2 {\cal D}_{\rm SSV}^\naive(m_1,m_2,\Mw)
          - {1\over4} g_2^2 {\cal D}_{\rm SSV}^\naive(m_2,m_2,\Mw)
   \cr &
          - {1\over8} {g_2^2\over\cos^2\thetaw}
                  {\cal D}_{\rm SSV}^\naive(m_1,m_2,\Mz)
          - {1\over8} {g_2^2\over\cos^2\thetaw}(1-2\sin^2\thetaw)^2
                 {\cal D}_{\rm SSV}^\naive(m_2,m_2,\Mz)
   \cr &
          - {1\over2} e^2 {\cal D}_{\rm SSV}^\naive(m_2,m_2,0)
          - {T^3\over96\pi} (g_2^2 \dMa + g_1^2 \dMb)
          + O(g^{9/2}T^4)
 & \AMSMdiags a \cr
   V^{\rm(b)}
        &= - \phi^2\biggl\{
             {1\over32} g_2^4 {\cal D}_{\rm SVV}(m_1,\Mw,\Mw)
   \cr&
           + {1\over64} {g_2^4\over\cos^4\thetaw}
                 [ {\cal D}_{\rm SVV}^\naive(m_1,\Mz,\Mz)
                   - 4\bar H_3(m_1,\Mz,\Mz) ]
   \cr&
           + {1\over16} g_2^2 e^2 \tan^2\thetaw
                 [ {\cal D}_{\rm SVV}^\naive(m_2,\Mw,\Mz)
                   - 4\bar H_3(m_2,\Mw,\Mz) ]
   \cr&
           + {1\over16} g_2^2 e^2
                 [ {\cal D}_{\rm SVV}^\naive(m_2,\Mw,0)
                   - 4\bar H_3(m_2,\Mw,0) ]
   \cr&
           + {1\over16}(g_2\cos\thetat + g_1\sin\thetat)^4
                 \bar H_3(m_1,\MLz,\MLz)
   \cr&
           + {1\over8} (g_2\cos\thetat + g_1\sin\thetat)^2
                       (g_2\sin\thetat - g_1\cos\thetat)^2
                 \bar H_3(m_1,\MLz,\MLg)
   \cr&
           + {1\over16}(g_2\sin\thetat - g_1\cos\thetat)^4
                 \bar H_3(m_1,\MLg,\MLg)
   \cr&
           + {1\over4} g_2^2 g_1^2 \sin^2\thetat \bar H_3(m_2,\MLw,\MLz)
           + {1\over4} g_2^2 g_1^2 \cos^2\thetat \bar H_3(m_2,\MLw,\MLg)
           \biggr\}
   \cr&
           + O(g^5 T^4)
 & \AMSMdiags b \cr
   V^{\rm(c)}
        &=
           - {1\over4} g_2^2 {\cal D}_{\rm SV}^\naive(m_2,\Mw)
           - {1\over8} g_2^2
                 [ {\cal D}_{\rm SV}^\naive(m_1,\Mw)
                 + {\cal D}_{\rm SV}^\naive(m_2,\Mw) ]
   \cr&
           - {1\over8} {g_2^2\over\cos^2\thetaw} (1-2\sin^2\thetaw)^2
                 {\cal D}_{\rm SV}^\naive(m_2,\Mz)
           - {1\over2} e^2 {\cal D}_{\rm SV}^\naive(m_2,0)
   \cr&
           - {1\over16}{g_2^2\over\cos^2\thetaw}
                 [ {\cal D}_{\rm SV}^\naive(m_1,\Mz)
                 + {\cal D}_{\rm SV}^\naive(m_2,\Mz) ]
           - {T^3\over96\pi} (g_2^2 \dMa + g_1^2 \dMb)
   \cr&
           - \left({3\over4}g_2^2+{1\over4}g_1^2\right)
                 (\tilde m_1 - m_1 + 3\tilde m_2 - 3m_2)
                 {T^3\over32\pi}
           + O(g^{9/2}T^4)
 & \AMSMdiags c \cr
   V^{\rm(d)}
        &= {T^2\over4(4\pi)^2} \left[
              g_1^2 \Mz^2\sin^2\thetaw \left({1\over12}+{10\nf\over9}\right)
              - 3 g_2^2\Mw^2\left({25\over12}-{2\nf\over3}\right)
           \right]
           \left({1\over\bar\eps} - {2\over3} \right)
   \cr&
           + O(g^5 T^4)
 & \AMSMdiags d \cr
   V^{\rm(e)}
        &= - {3\phi^2 T^2\over128(4\pi)^2}
                (3g_2^4-6g_2^2g_1^2-g_1^4+12g_2^2\gy^2+4g_1^2\gy^2)
                \left({1\over\bar\eps}-{2\over3}\right)
           + O(g^{9/2}T^4)
 & \AMSMdiags e \cr
   V^{\rm(f)}
        &= O(g^5T^4)
 & \AMSMdiags f \cr
   V^{\rm(g)}
        &= \left( {9\over8}g_2^4 + {3\over4}g_2^2g_1^2
                    + {3\over8}g_1^4 - 6\gy^4 \right)
             {\phi^2 T^2 \over 8(4\pi)^2} {1\over\bar\eps}
        + O(g^{9/2}T^4)
 & \AMSMdiags g \cr
   V^{\rm(h)}
        &= g_2^2 \left({5\over6}+{\nf\over3}\right)
              (2\MLw + \cos^2\thetat\MLz + \sin^2\thetat\MLg) {T^3\over8\pi}
   \cr&
           + g_1^2 \left({1\over6}+{5\nf\over9}\right)
              (\sin^2\thetat\MLz + \cos^2\thetat\MLg) {T^3\over8\pi}
   \cr&
           + \left({3\over4}g_2^2+{1\over4}g_1^2+\gy^2\right)
              (\tilde m_1 + 3\tilde m_2) {T^3\over32\pi}
           + O(g^{9/2}T^4)
 & \AMSMdiags h \cr
   V^{\rm(i)}
        &= - 4\gs^2 {\cal D}_{\rm FFV}^\naive(\mf,\mf,0)
   \cr&
           - {3\over16}{g_2^2\over\cos^2\thetaw}
               \left[\left(1-{8\over3}\sin^2\thetaw\right)^2 + 1\right]
               [ {\cal D}_{\rm FFV}^\naive(\mf,\mf,\Mz)
                  - {\cal D}_{\rm FFV}^\naive(0,0,\Mz) ]
   \cr&
           - {3\over2} g_2^2
               [ {\cal D}_{\rm FFV}^\naive(\mf,0,\Mw)
                  - {\cal D}_{\rm FFV}^\naive(0,0,\Mw) ]
           - 2\nf g_2^2 {\cal D}_{\rm FFV}^\naive(0,0,\Mw)
   \cr&
           - \nf g_2^2 \left( \cos^2\thetaw +
                           {5\over3}{\sin^4\thetaw\over\cos^2\thetaw} \right)
               {\cal D}_{\rm FFV}^\naive(0,0,\Mz)
           - {4\over3} e^2 {\cal D}_{\rm FFV}^\naive(\mf,\mf,0)
   \cr&
           - {\nf T^3\over24\pi} (g_2^2 \dMa + {5\over3}g_1^2 \dMb)
           + O(g^5 T^4)
 & \AMSMdiags i \cr
   V^{\rm(j)}
      &= - {3\over4} \gy^2 {\cal D}_{\rm FFS}(\mf,\mf,m_1)
         - {3\over4} \gy^2 {\cal D}_{\rm FFS}(\mf,\mf,m_2)
         - {3\over2} \gy^2 {\cal D}_{\rm FFS}(\mf,0,m_2)
 & \AMSMdiags j \cr
   V^{\rm(k)}
        &= {3\gy^2\mf^2 T^2\over8(4\pi)^2}
              \left({1\over\bar\eps} - 2\ln2 \right)
         + O(g^6 T^4)
 & \AMSMdiags k \cr
   V^{\rm(l)}
        &= - \left(\gs^2-{1\over12}g_1^2\right)
              {\sqrt2\gy\mf\phi T^2 \over (4\pi)^2}
              \left({1\over\bar\eps} - 2\ln2 \right)
         + O(g^6 T^4)
 & \AMSMdiags l \cr
   V^{\rm(m)}
        &= - {1\over2}g_2^2\cos^2\thetaw
               {\cal D}_{\rm VVV}^\naive(\Mw,\Mw,\Mz)
           - {1\over2} e^2 {\cal D}_{\rm VVV}^\naive(\Mw,\Mw,0)
   \cr&
           - {1\over2}g_2^2\cos^2\thetaw \left[
                 {\cal D}_{\rm LLT}(\MLw,\MLw,\Mz)
                 - {\cal D}_{\rm LLT}(\Mw,\Mw,\Mz) \right]
   \cr&
           - {1\over2} e^2 \left[
                 {\cal D}_{\rm LLT}(\MLw,\MLw,0)
                 - {\cal D}_{\rm LLT}(\Mw,\Mw,0) \right]
   \cr&
           - g_2^2\cos^2\thetat {\cal D}_{\rm LLT}(\MLw,\MLz,\Mw)
           + g_2^2\cos^2\thetaw {\cal D}_{\rm LLT}(\Mw,\Mz,\Mw)
   \cr&
           - g_2^2\sin^2\thetat {\cal D}_{\rm LLT}(\MLw,\MLg,\Mw)
           + g_2^2\sin^2\thetaw {\cal D}_{\rm LLT}(\MLw,0,\Mw)
   \cr&
           - {g_2^2 T^3\over16\pi} \dMa
           + O(g^5 T^4)
 & \AMSMdiags m \cr
   V^{\rm(n)}
        &= - 2 g_2^2 {\cal D}_{\eta\eta\rm V}^\naive(\Mw)
           - g_2^2\cos^2\thetaw {\cal D}_{\eta\eta\rm V}^\naive(\Mz)
           + {g_2^2 T^3\over96\pi} \dMa
           + O(g^5 T^4)
 & \AMSMdiags n \cr
   V^{\rm(o)}
        &= - {1\over4}g_2^2 {\cal D}_{\rm VV}^\naive(\Mw,\Mw)
           - {1\over2}g_2^2\cos^2\thetaw {\cal D}_{\rm VV}^\naive(\Mw,\Mz)
           - {1\over2}e^2 {\cal D}_{\rm VV}^\naive(\Mw,0)
   \cr&
           - {g_2^2 T^3\over32\pi} \dMa
           + {2g_2^2 T^2\over(4\pi)^2} \Mw\dMa
           + O(g^5 T^4)
 & \AMSMdiags o \cr
 }
$$

\subsec{Results for resummed graphs in terms of unresummed graphs}
\eqn\dssvR{
  \eqalign{
    {\cal D}_{\rm SSV}(m_1,m_2,M)
        =& {\cal D}_{\rm SSV}^\naive(m_1,m_2,M)
         + {1\over24\pi} (\ML-M) T^3
         + O(g^{5/2} T^4)
  }
}
\eqn\dsvvR{
    {\cal D}_{\rm SVV}(m,M_1,M_2)
        = {\cal D}_{\rm SVV}^\naive(m,M_1,M_2)
        + 4 \bar H_3(\tilde m,\MLa,\MLb) - 4 \bar H_3(m,M_1,M_2)
        + O(g T^2)
}
\eqn\dsvR{
    {\cal D}_{\rm SV}(m,M)
        = {\cal D}_{\rm SV}^\naive(m,M)
        + {1\over24\pi} (\ML - M) T^3
        + {1\over8\pi} (\tilde m - m) T^3
        + O(g^{5/2} T^4)
}
\eqn\dffvR{
    {\cal D}_{\rm FFV}(\mfa,\mfb,M)
        = {\cal D}_{\rm FFV}^\naive(\mfa,\mfb,M)
        + {1\over24\pi} (\ML - M) T^3
        + O(g^3 T^4)
}
\eqn\dffsR{
    {\cal D}_{\rm FFS}(\mfa,\mfb,m)
        = {\cal D}_{\rm FFS}^\naive(\mfa,\mfb,m)
        + {1\over24\pi} (\tilde m - m) T^3
        + O(g^{7/2} T^4)
}
\eqn\dvvvR{
  \eqalign{
    {\cal D}_{\rm VVV}(M_1,M_2,M_3)
  &
        = {\cal D}_{\rm VVV}^\naive(M_1,M_2,M_3)
        + {1\over8\pi} (\MLa+\MLb+\MLc-M_1-M_2-M_3) T^3
  \cr&
        + {1\over2}\sum_{\rm perms.}
             [ {\cal D}_{\rm LLT}(\MLa,\MLb,\MLc)
              - {\cal D}_{\rm LLT}(M_1,M_2,M_3) ]
        + O(g^3T^4)
  }
}
The sum above is over all six permutations of $(M_1,M_2,M_3)$.
\eqn\dllt{
  \eqalign{
   {\cal D}_{\rm LLT}(M_1,M_2,&M_3)
       = {T^2\over(4\pi)^2} \left[
            (M_1+M_2)M_3 - M_1M_2 + {(M_1+M_2)(M_1-M_2)^2\over M_3} \right]
  \cr &
       + (M_3^2 - 2M_1^2 - 2M_2^2) \bar H_3(M_1,M_2,M_3)
  \cr &
       + {(M_1^2-M_2^2)^2\over M_3^2}
                [ \bar H_3(M_1,M_2,M_3) - \bar H_3(M_1,M_2,0) ]
  }
}
\eqn\dnnvR{
   {\cal D}_{\eta\eta \rm V}(M)
       = {\cal D}_{\eta\eta \rm V}^\naive(M)
       - {1\over96\pi} (\ML-M) T^3
       + O(g^3 T^4)
}
\eqn\dvvR{
  \eqalign{
    {\cal D}_{\rm VV}(M_1,M_2)
       = & {\cal D}_{\rm VV}^\naive(M_1,M_2)
       + {1\over16\pi} (\MLa+\MLb-M_1-M_2) T^3
  \cr&
       - \bigl[(\MLa-M_1)M_2 + M_1(\MLb-M_2)\bigr] {4T^2\over(4\pi)^2}
       + O(g^3 T^4)
  }
}

\subsec{Exact results for unresummed graphs in terms of
        $I$, $\bar H$ and $L$}

This section is included only for completeness.  The expansions for
the ${\cal D}'s$ that give our final result for the effective potential
through $O(g^4 T^4)$ are given in the next section.  We haven't bothered
to keep track of the $\bar\Hff$ terms in the fermion contributions
below because these do not contribute to the potential at $O(g^4 T^4)$.
In the case of ${\cal D}_{\rm FFS}$, these terms in fact depend on
whether the S is a scalar or a pseudo-scalar---a distinction we are
otherwise able to ignore.

\eqn\dssv{
  \eqalign{
    {\cal D}_{\rm SSV}^\naive(m_1,m_2,&M)
       = I(M) [I(m_1)+I(m_2)] - I(m_1)I(m_2)
  \cr &
       + (M^2-2m_1^2-2m_2^2) \bar H(m_1,m_2,M)
  \cr &
       + {(m_1^2-m_2^2)\over M^2} [I(M)-I(0)] [I(m_1)-I(m_2)]
  \cr &
       + {(m_1^2-m_2^2)^2\over M^2} [\bar H(m_1,m_2,M) - \bar H(m_1,m_2,0)]
  }
}
\eqn\dsvv{
  \eqalign{
    {\cal D}_{\rm SVV}^\naive(m,M_1,&M_2)
    = \biggl\{
       {1\over M_1^2} [I(M_1)-I(0)][I(M_2)-I(m)]
  \cr &
       - {m^2\over 2M_1^2 M_2^2} [I(M_1)-I(0)][I(M_2)-I(0)]
       + (5-4\eps) \bar H(m,M_1,M_2)
  \cr &
       + {m^4\over 2M_1^2M_2^2} [\bar H(m,M_1,M_2) - 2\bar H(m,M_1,0) +
             \bar H(m,0,0) ]
  \cr &
       + {(M_1^2-2m^2)\over M_2^2} [\bar H(m,M_1,M_2) - \bar H(m,M_1,0)]
    \biggr\}
       ~~+~~ \{ M_1 \leftrightarrow M_2 \}
  }
}
\eqn\dsv{
    {\cal D}_{\rm SV}^\naive(m,M)
       = - 2 (3-2\eps) I(m) I(M)
}
\eqn\dffv{
  \eqalign{
    {\cal D}_{\rm FFV}^\naive(\mfa,\mfb,&M)
       = 2(1-\eps) [\If(\mfa)+\If(\mfb)] I(M)
       - 2(1-\eps) \If(\mfa) \If(\mfb)
  \cr&
       - {(\mfa^2-\mfb^2)\over M^2} [\If(\mfa)-\If(\mfb)] [I(M)-I(0)]
       + O(g^2 T^2 \bar\Hff)
  }
}
\eqn\dffs{
    {\cal D}_{\rm FFS}^\naive(\mfa,\mfb,m)
       = 2 [\If(\mfa)+\If(\mfb)] I(m)
       - 2 \If(\mfa) \If(\mfb)
       + O(g^2 T^2 \bar\Hff)
}
\eqn\dvvv{
  \eqalign{
    {\cal D}_{\rm VVV}^\naive(M_1,M_2,&M_3)
    = {1\over2} \sum_{\rm perms.} \biggl\{
      {1\over4} {M_1^6\over M_2^2 M_3^2} \bar H(M_1,0,0)
      - {1\over4} {M_1^4\over M_2^2 M_3^2} [I(0)]^2
      - L(M_1,M_2)
  \cr &
      + \left[ (-4+4\eps){M_1^4\over M_3^2}
             + \left({9\over2}-4\eps\right){M_1^2 M_2^2\over M_3^2}
             - {1\over2}{M_1^6\over M_2^2 M_3^2}
            \right]\bar H(M_1,M_2,0)
  \cr &
      + \left[ \left({7\over2}-2\eps\right)
             + \left({9\over2}-4\eps\right)
                      \left({M_1^2\over M_2^2}-{M_3^2\over M_1^2}\right)
             - {1\over4}{M_3^4\over M_1^2 M_2^2}
            \right] I(M_1) I(M_2)
  \cr &
      + \left[ \left({9\over2}-4\eps\right)
                      \left({M_2^2\over M_3^2}-{M_1^2\over M_2^2}\right)
             + {1\over2} {M_2^4\over M_1^2 M_3^2}
            \right] I(M_1) I(0)
  \cr &
      + \left[ (-8+6\eps)M_1^2 + (4-4\eps){M_1^4\over M_3^2}
             + \left(-{9\over2}+4\eps\right){M_1^2 M_2^2\over M_3^2}
             + {1\over4} {M_1^6\over M_2^2 M_3^2}
            \right]
  \cr & \qquad\qquad\qquad \times
        \bar H(M_1,M_2,M_3)
    \biggr\}
  }
}
The sum above is over all six permutations of $(M_1,M_2,M_3)$.
\eqn\dnnv{
   {\cal D}_{\eta\eta \rm V}(M)
       = - {1\over2} I(M) I(0) - {1\over4} M^2 \bar H(M,0,0)
}
\eqn\dvv{
    {\cal D}_{\rm VV}^\naive(M_1,M_2)
    = 2L(M_1,M_2) + (-14+20\eps) I(M_1) I(M_2)
}

\subsec{Expansion of results for unresummed graphs}

\eqn\dssvX{
  \eqalign{
    {\cal D}_{\rm SSV}^\naive(m_1,m_2,M)
       =& - {M T^3 \over 24\pi}
       + {M^2 T^2 \over (4\pi)^2} \left[
           {1\over12}{1\over\bar\eps}
           \!+\! {1\over12}\ln\left(\mub^2\over T^2\right)
           + \ln\left(3T\over M\right)
           + {1\over2} + {1\over3}\cb \!-\! {1\over4}\cH
          \right]
  \cr &
       + O(g^{5/2} T^4) .
  }
}
\eqn\dsvvX{
  \eqalign{
    {\cal D}_{\rm SVV}^\naive(m,M_1,&M_2)
       = {T^2 \over (4\pi)^2} \biggl[
           {5\over2}{1\over\bar\eps}
           + {5\over2}\ln\left(\mub^2\over T^2\right)
           + 3
           - {5\over2} \cH
           + {M_1\over M_2}
  \cr&
           + {M_2\over M_1}
           + \left( 10 + {M_1^2\over M_2^2} + {M_2^2\over M_1^2} \right)
                         \ln\left(3T\over M_1+M_2\right)
  \cr&
           - {M_1^2\over M_2^2} \ln\left(3T\over M_1\right)
           - {M_2^2\over M_1^2} \ln\left(3T\over M_2\right)
         \biggr]
       + O(g^{1/2}T^2) .
  }
}
\eqn\dsvX{
    {\cal D}_{\rm SV}^\naive(m,M)
       = {M^2 T^2\over (4\pi)^2} \left[
           {1\over2} {1\over\bar\eps}
           + {1\over2} \ln\left(\mub^2\over T^2\right)
           - {1\over3} - \cb \right]
       + {M T^3\over 8\pi}
       + {mT^3\over 8\pi}
       + O(g^{7/2}T^4)
}
\eqn\dffvX{
  \eqalign{
    {\cal D}_{\rm FFV}^\naive(\mfa,\mfb,&M)
       = {(\mfa^2+\mfb^2) T^2 \over (4\pi)^2} \left[
           - {1\over4}{1\over\bar\eps}
           - {1\over4} \ln\left(\mub^2\over T^2\right)
           + {1\over4} + {1\over2}\cf + {1\over6}\ln2 \right]
  \cr &
       + {M^2 T^2\over(4\pi)^2} \left[
           {1\over6}{1\over\bar\eps}
           + {1\over6} \ln\left(\mub^2\over T^2\right)
           - {1\over6} - {1\over3}\cb - {1\over3}\ln 2 \right]
       + {MT^3\over24\pi}
       + O(g^3 T^4)
  }
}
\eqn\dffsX{
  \eqalign{
    {\cal D}_{\rm FFS}^\naive(\mfa,\mfb,m)
       =& {(\mfa^2+\mfb^2) T^2 \over (4\pi)^2} \left[
           - {1\over4}{1\over\bar\eps}
           - {1\over4} \ln\left(\mub^2\over T^2\right)
           + {1\over2}\cf + {1\over6}\ln2 \right]
       + {mT^3\over24\pi}
  \cr&
       + O(g^{7/2}T^4)
  }
}
\eqn\dvvvX{
  \eqalign{
    {\cal D}_{\rm VVV}^\naive(M,M,&M_3)
       = {T^2\over(4\pi)^2} \biggl[
           (2M^2+M_3^2) \left(
              - {61\over24}{1\over\bar\eps}
              - {61\over24}\ln{\mub^2\over T^2}
              + {13\over8}
              + {13\over12}\cb
              + 2\cH            \right)
  \cr&
           + {11\over6}MM_3 - {25\over3}M_3^2 + {9M_3^3\over2M}
           - {M^3\over2M_3} - {M_3^4\over4M^2}
  \cr&
           + \left( - 4M^2 + 9M_3^2 - {M^4\over2M_3^2} - {4M_3^4\over M^2}
                    - {M_3^6\over2M^4} \right) \ln\left(3T\over M+M_3\right)
  \cr&
           + \left( - 12M^2 - 17 M_3^2 + {4M_3^4\over M^2}
                    + {M_3^6\over4M^4} \right) \ln\left(3T\over2M+M_3\right)
  \cr&
           + {M^4\over2M_3^2}\ln\left(3T\over M\right)
           + {M_3^6\over4M^4}\ln\left(3T\over M_3\right)
       \biggr]
       - (2M+M_3){T^3\over8\pi}
       + O(g^3 T^4)
  }
}
\eqn\dnnvX{
  \eqalign{
    {\cal D}_{\eta\eta\rm V}^\naive(M)
       =  & {M^2T^2\over(4\pi)^2} \left[
             - {1\over48}{1\over\bar\eps}
             - {1\over48}\ln{\mub^2\over T^2}
             - {1\over8} - {1\over12}\cb + {1\over16}\cH
             - {1\over4}\ln\left(3T\over M\right)
           \right]
  \cr&
           + {M T^3\over 96\pi}
           + O(g^3 T^4)
  }
}
\eqn\dvvX{
  \eqalign{
    {\cal D}_{\rm VV}^\naive(M_1,M_2)
       =  {T^2\over(4\pi)^2} \biggl[
             (M_1^2+M_2^2) & \left(
               {9\over8}{1\over\bar\eps}
               + {9\over8}\ln{\mub^2\over T^2}
               - {13\over8} - {9\over4} \cb \right)
             - {40\over3}M_1M_2 \biggr]
  \cr&
           + {13\over48\pi}(M_1+M_2)T^3
           + O(g^3 T^4)
  }
}

\subsec{Some useful limits}

\eqn\dvvvL{
  \eqalign{
    {\cal D}_{\rm VVV}^\naive(M,M,0)
       = {M^2T^2\over(4\pi)^2} \biggl[
   &
             - {61\over12}{1\over\bar\eps}
             - {61\over12}\ln{\mub^2\over T^2}
             + 3 + {13\over6}\cb + 4\cH
             + 12\ln2
   \cr&
             - 16\ln\left(3T\over M\right)
          \biggr]
          + O(g^3 T^4)
  }
}
\eqn\dsvvL{
    {\cal D}_{\rm SVV}^\naive(m,M,0)
       = {T^2\over(4\pi)^2} \biggl[
           {5\over2}{1\over\bar\eps}
           + {5\over2}\ln{\mub^2\over T^2}
           + {7\over2} - {5\over2}\cH
           + 10\ln\left(3T\over M\right)
       \biggr]
       + O(g^{1/2} T^2)
}
\eqn\dlltL{
    {\cal D}_{\rm LLT}(M_1,M_2,0)
      = {T^2\over(4\pi)^2} \left({1\over2}M_1^2 - 2M_1M_2
                                   + {1\over2}M_2^2\right)
        - 2(M_1^2+M_2^2) \bar H_3(M_1,M_2,0)
}

\appendix{B}{Derivation of Some High-Temperature Expansions}

\subsec{Expansion of $\bar\Hff$}

In this appendix, we shall derive the leading $O(T^2)$ term in the
high-temperature expansion of $\bar\Hff(\mfa,\mfb;m)$, defined by
\Hfdef.  Because two of the three Euclidean momenta $P$, $Q$, $P+Q$ are
fermionic, $\bar\Hff$ does not diverge in the infra-red if we set
all the masses to zero.  So the leading $O(T^2)$ term is simply
\eqn\Hmassless{
   \bar\Hff(\mfa,\mfb;m)
         = \mu^{4\eps} \sumintf_P \sumintf_Q {1 \over P^2 Q^2 (P+Q)^2}
                         + O(mT) .
}
The result found in this appendix is that
\eqn\wow{
   \sumintf_P \sumintf_Q {1 \over P^2 Q^2 (P+Q)^2} = O(\eps)
}
in dimensional regularization and so vanishes when $\eps\rightarrow 0$.

We begin by rewriting the integral in \wow\ as
\eqn\wowb{
  \eqalign{
     T^2
     \sum_{\scriptstyle {\rm odd}  \atop \scriptstyle n}
     \sum_{\scriptstyle {\rm odd}  \atop \scriptstyle j}
  &
     \sum_{\scriptstyle {\rm even \vphantom{d}} \atop \scriptstyle k}
     \delta_{n+j+k}
     \int {\d^{3-2\eps}p\over(2\pi)^{3-2\eps}}
          {\d^{3-2\eps}q\over(2\pi)^{3-2\eps}}
          {\d^{3-2\eps}r\over(2\pi)^{3-2\eps}}
     \delta(p+q+r)
  \cr & \times
     {1 \over [(n \pi T)^2+p^2] [(j \pi T)^2+q^2]
              [ (k \pi T)^2 + r^2 ] }.
  }
}
Using the symmetry of this expression, one may replace
\eqn\wowc{
   \sum_{\rm odd} \sum_{\rm odd} \sum_{\rm even \vphantom{d}}
   \rightarrow
   {1\over3} \left(
      \sum_{\rm any} \sum_{\rm any} \sum_{\rm any}
      - \sum_{\rm even} \sum_{\rm even} \sum_{\rm even}
   \right) ,
}
where we will restrict the triple sum on the right-hand side to
exclude the case $(n,j,l)=(0,0,0)$ where all the frequencies are
simultaneously zero.
By scaling all three-momenta by a factor of 2 in the first term
on the right-hand side of \wowc, we obtain
\eqn\wowd{
   \sum_{\rm odd} \sum_{\rm odd} \sum_{\rm even \vphantom{d}}
   \rightarrow
   {(2^{4\eps}-1)\over3} \sum_{\rm even} \sum_{\rm even} \sum_{\rm even} .
}
So we may relate $\bar\Hff$ to something resembling the bosonic
function $\bar H$:
\eqn\wowd{
   \sumintf_P \sumintf_Q {1\over P^2 Q^2 (P+Q)^2}
   = \eps [4\ln2 + O(\eps)]
     \sumint_P \sumint_Q {1-\delta_{p_0}\delta_{q_0}
                      \over P^2 Q^2 (P+Q)^2 } .
}
The result will be non-zero when $\eps\rightarrow0$ only if the
bosonic integrals on the right-hand side give a divergent contribution
of order $1/\eps$.  To relate this result more directly to the
bosonic function $H(m)$, we need to temporarily put in an infrared
cut-off $m$:
\eqn\wowe{
  \eqalign{
  &
     \sumint_P \sumint_Q {1-\delta_{p_0}\delta_{q_0}
                      \over P^2 Q^2 (P+Q)^2 } =
  \cr & \qquad
     \lim_{m\rightarrow 0}
       \left\{ H(m)
            - T^2 \int {\d^{3-2\eps}p\over(2\pi)^{3-2\eps}}
                       {\d^{3-2\eps}q\over(2\pi)^{3-2\eps}}
                       {1\over(p^2+m^2)(q^2+m^2)[(p+q)^2+m^2]}
       \right\}
  }
}
It is straightforward to compute the $1/\eps$ term of the
above three-dimensional integrals using standard techniques
for loop integrals.  One finds that it exactly cancels the
$1/\eps$ piece of $H(m)$ shown in \eqH, from which \wow\
then follows.

\subsec{Expansion of $L(m_1,m_2)$}

Working from the definition \defL\ of $L(m_1,m_2)$,
\eqn\deriveL{
   \eqalign{
      L(m_1,m_2)
   &
      =
      \sumint {p_0^2 \over P^2 (P^2+m_1^2)}
      \sumint {q_0^2 \over Q^2 (Q^2+m_2^2)}
      + \sum_{i,j=1}^{3-2\eps}
      \sumint {p_i p_j \over P^2 (P^2+m_1^2)}
      \sumint {q_i q_j \over Q^2 (Q^2+m_2^2)}
   \cr&
      =
      \sumint {p_0^2 \over P^2 (P^2+m_1^2)}
      \sumint {q_0^2 \over Q^2 (Q^2+m_2^2)}
   \cr& \qquad
      + {1\over(3-2\eps)}
      \left[ I(m_1) - \sumint {p_0^2 \over P^2 (P^2+m_1^2)} \right]
      \left[ I(m_2) - \sumint {p_0^2 \over P^2 (P^2+m_2^2)} \right] .
   }
}
An expansion for the remaining integral may easily be found by the
method of \deriveI:
\eqn\eqIzero{
  \eqalign{
     \sumint {p_0^2 \over P^2 (P^2+m^2)} =
  &
       - {1\over24}T^2
       - {1\over64\pi^2} m^2
         \left[ {1\over\eps} + \ln\left(\mub^2\over T^2\right)
                - 2\cb + 2 \right]
       + O(m^4/T^2)
  \cr &
       - \eps (\ceps - 2) {1\over24} T^2
       + O(\eps m^2) .
  }
}
The high-temperature expansion of $L(m_1,m_2)$ is then
\eqn\eqL{
  \eqalign{
     L(m_1,m_2) =
     \const
  &
     - {(m_1+m_2)T^3\over24\cdot4\pi}
     - (m_1^2+m_2^2) {T^2\over48(4\pi)^2} \left(
         {1\over\eps} + \ceps + \ln{\mub^2\over T^2} - 2\cb - 1 \right)
  \cr &
     + {m_1m_2 T^2\over3(4\pi)^2}
     + O(m^3 T) + O(\eps)
  }
}

\appendix{C}{Criticism of the Super-Daisy Approximation}

\PSfig\figz{
  The one-loop approximation to Dyson's equation for the self-energy.
}{3cm}{fig28.psx}{90}

\PSfig\figA{
  A generic super-daisy diagram in the pure scalar theory.
}{15.5cm}{fig29.psx}{60}

A previous attempt to compute corrections to the ring-improved
one-loop potential in ref.~\boyd\ has relied on the super-daisy approximation
to simplify the calculation.  In this appendix, we shall explain
why the super-daisy approximation does not correctly reproduce
the leading corrections to the one-loop potential.
The super-daisy approximation is an approximation to Dyson's
equations where the effective masses of particles are derived at
one loop (to any desired order in $M/T$) from diagrams like
Figs.~\xfig\figa\ and \xfig\fige, and then those masses are used
self-consistently
to improve the propagators used to derive the effective masses
in the first place.  The result is a set of coupled equations
which may be solved for the effective masses.  In pure scalar
theory, for example, the equation is given schematically
by \figz\
(ignoring renormalization and counter-terms):
\eqn\SDdyson{
  \meff^2 \approx m^2 + {1\over2} \lambda \sumint_P {1\over P^2+\meff^2}
          - {1\over2} \lambda^2 \phi^2 \sumint_P {1\over(P^2+\meff^2)^2} .
}
Computing the
one-loop potential with these masses is then equivalent to
some approximation to the set of super-daisy diagrams, such as
shown in \figA.

Super-daisy equations such as \SDdyson, however, do not give
adequate approximations to these graphs.  \SDdyson\ should
really read
\eqn\SDdysonb{
  \Pi(Q) \approx m^2 + {1\over2} \lambda \sumint_P {1\over P^2+\Pi(P)}
          - {1\over2} \lambda^2 \phi^2 \sumint_P
               {1\over [P^2+\Pi(P)] [(P+Q)^2+\Pi(P+Q)]} ,
}
where $\Pi(P)$ is the one-loop self-energy.
The approximation made in \SDdyson\ was to replace $\Pi(P)$ by
$\Pi(0)$ inside loops.  Thus, \figA\ has been approximated by
first computing the smallest (outer-most) loops in the approximation
that no momenta flows into them, then computing the next smallest
loops in the same approximation, then the next, and so forth.
At each stage, the approximation is valid only if the momentum
flowing through a loop is small compared to the momenta of
all the smaller loops that decorate it.  And therein lies the rub.
Once quadratically divergent pieces have been accounted for
(say, by normal ring resummation as opposed to super-daisy
resummation), then most loops are dominated by momenta of
order $M$ and there is no hierarchy of momenta in a graph.
So, for example, the $O(g^2 M^2 T^2)$ contributions of
\figu m cannot be correctly computed in an approximation that replaces
one loop by $\Pi(0)$.

\PSfig\figB{
  A disastrous term in the super-daisy resummation of the
  one-loop vector diagram.
}{5.5cm}{fig30.psx}{80}

\PSfig\figC{
  The tadpole version of \figB, which is still problematical.
}{7cm}{fig31.psx}{80}

The problem can be even more egregious if care isn't taken to
avoid mishandling hard ($P \sim T$) thermal loops.
Consider the particular contribution to the super-daisy improved
one-loop potential shown on the right-hand side of \figB.
The super-daisy approximation instructs us to first evaluate
the small loop assuming $Q = 0$.  This small loop is then dominated
by loop momenta $P$ of order $M$.  When we
replace the result $\Pi(P)$ of the small loop by an effective mass insertion
$\Pi(0)$, the large loop becomes quadratically divergent and
so is dominated by momenta $Q \sim T$.  But this is exactly the
{\it opposite} hierarchy of momenta from the one needed to justify
the super-daisy approximation.  Ref.~\boyd\ sidesteps this disaster
by computing the derivative of the effective potential rather
than the effective potential itself.  In the tadpole graph of \figC,
the large loop never has momentum $T$ when it is decorated
by smaller loops.  But this does {\it not} save the super-daisy approximation.
The large loop is now dominated by momenta $Q \sim M$, and the
$O(MT)$ corrections from the small loop are also dominated by
momenta $P \sim M$; the necessary hierarchy of momenta is still
absent.

As discussed in section \DoubleDaisySec,
there are a few diagrams where there
{\it is} a hierarchy of scales
when the scalar mass is much smaller than the vector mass.
The resummation discussed in that section does not benefit from
the full super-daisy approximation of \SDdyson\ and was in any case
irrelevant to our final results for the leading corrections to the
one-loop potential.

In section \secShift, we discussed how the most important terms for
shifting the VEV from its one-loop value are those that depend on
logarithms of masses.  The super-daisy approximation will never
generate such terms because it always reduces multi-loop diagrams
to a hierarchy of one-loop diagrams (with some number of mass
insertions) and the result $I(M)$ for a one-loop diagram does
not contain any $\ln M$ terms.  [See \eqI.]  Such terms do appear
in many of the expressions in ref.~\boyd, but this is because there
$I(M)$ is routinely split into separate zero-temperature and
temperature-dependent pieces.  Though each piece depends on $\ln M$,
the sum does not.  Ref.~\boyd\ does consider one diagram in
addition to the super-daisy one-loop tadpole, and this one diagram
will indeed generate some $\ln M$ terms.  However the other $\ln M$ terms
from all those diagrams subsumed by the super-daisy one-loop tadpole
(such as the derivative of \figu m, for example) are lost.

\listrefs
\if\preprint0 \listfigs \fi

\end